\documentclass[aps,pre,showpacs,twocolumn]{revtex4}

\usepackage{epsfig}
\usepackage{float}
\usepackage{graphicx}
\usepackage{graphics}
\usepackage{color}

\begin{document}


\title{Finite size scaling for a first order transition where a continuous symmetry is broken:  The spin-flop transition in the $3D$ XXZ Heisenberg antiferromagnet}


\author{Jiahao Xu$^1$}
\email[]{jiahaoxu@uga.edu}
\author{Shan-Ho Tsai$^{1,2}$}
\email[]{shtsai@uga.edu}
\author{D. P. Landau$^{1,3}$}
\email[]{dlandau@hal.physast.uga.edu}
\author{K. Binder$^3$}
\email[]{kurt.binder@uni-mainz.de}
\affiliation{{\rm 1}. Center for Simulational Physics, University of Georgia, Athens, GA 30602 USA\\ {\rm 2}. Georgia Advanced Computing Resource Center, Enterprise Information Technology Services, University of Georgia, Athens, GA 30602 USA \\ {\rm 3}. Institut f\"ur Physik, Johannes Gutenberg Universit\"at Mainz, 55099 Mainz, Germany}


\date{\today}

\begin{abstract}
Finite size scaling for a first order phase transition where a continuous symmetry is broken is developed using an approximation of Gaussian probability distributions with a phenomenological ``degeneracy'' factor included.  Predictions are compared with data from Monte Carlo simulations of the three-dimensional, XXZ Heisenberg 
antiferromagnet in a field in order to study the finite size behavior on a $L \times L \times L$ simple cubic lattice for the first order ``spin-flop'' transition between the Ising-like antiferromagnetic state and the canted, XY-like state.  Our theory predicts that for large linear dimension $L$ the field dependence of all moments of the order parameters as well as the fourth-order cumulants exhibit universal intersections.  Corrections to leading order should scale as the inverse volume.  The values of these intersections at the spin-flop transition point can be expressed in terms of a factor $q$ that characterizes the relative degeneracy of the ordered phases.  Our theory yields $q=\pi$, and we present numerical evidence that is compatible with this prediction.
The agreement between the theory and simulation implies a heretofore unknown universality can be invoked for first order phase transitions.
\end{abstract}


\maketitle

\section{Introduction}
Finite size scaling at both first order and second order transitions between phases with discrete numbers of states is now relatively well established and extremely successful at describing phase transition behavior in the thermodynamic limit from Monte Carlo data produced for finite size systems ~\cite{MEF1,VP,DPL,BL1,BK,CLB,MCbook}. The simplest case of a first order transition, namely the first order, field driven transition in the 2-dimensional Ising model below its critical point, was studied using Monte Carlo simulations by Binder and Landau~\cite{BL1}; and to a good approximation the probability distribution of the order parameter at the first order transition could be described by the sum of the Gaussians representing the two coexisting states in the finite system.  For the temperature driven first-order transition in the $q$-state Potts model a similar theoretical development could be used, but a factor of ``$q$'' needed to be included in the Gaussian representing the $q$-fold degenerate ordered state~\cite{BK}.  Monte Carlo simulations of the $q=10$ Potts model on $L \times L$ lattices verified this finite size behavior~\cite{CLB}.

\begin{figure}
\centering
\includegraphics[width=0.5\hsize]{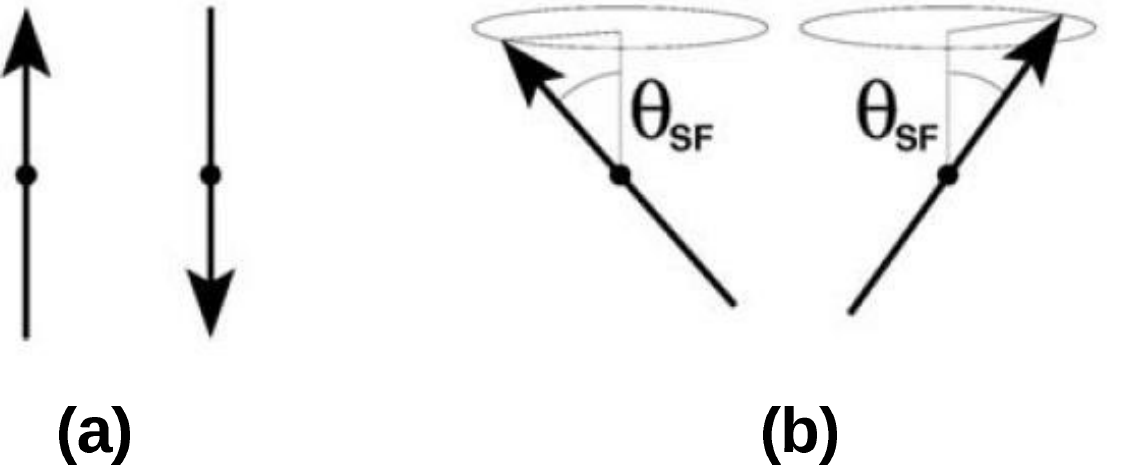}
\caption{\label{config}  Spin configurations for different phases in the anisotropic Heisenberg model in an applied field. Illustrated are spin configurations of the two sublattices in the (a) antiferromagnetic (AF); and (b) spin-flop (SF) phases. $\theta_{SF}$ is the angle that the spins make with respect to the applied field.}
\end{figure}

In the case of a first order phase transition involving the breaking of a continuous symmetry, however, there are neither good data from simulations nor theoretical predictions regarding the finite size behavior.  A good ``testing ground'' system is thus needed to help provide an understanding of this case, and we believe that the
uniaxially anisotropic, three-dimensional ($3D$) Heisenberg antiferromagnet in an external field, $H$, is exactly such a candidate model.  It has attracted substantial interest for a number of
 decades, largely to clarify the phase diagram and ordered structures for this model and to identify the nature of the multicritical point \cite{MEF,DRJMME,KNF,DLKB,MHJ,PCAPEV,RFYH,GBWS,WS,HTL}.  Although different scenarios have been proposed, we now believe that the phase diagram of this model contains a low temperature, low field antiferromagnetic (AF) phase in which the spins point in opposite directions along the axis given by the anisotropy
as shown in Fig. \ref{config}(a) and a spin-flop (SF) phase in which the spins are tilted with continuous rotational symmetry about the field direction  (see Fig. \ref{config}(b))
at low $T$ and higher $H$.  A paramagnetic (P) phase with no long range order exists at high $T$ and/or at high $H$. 

\begin{figure}
\centering
\includegraphics[angle = 0,width=0.85\hsize]{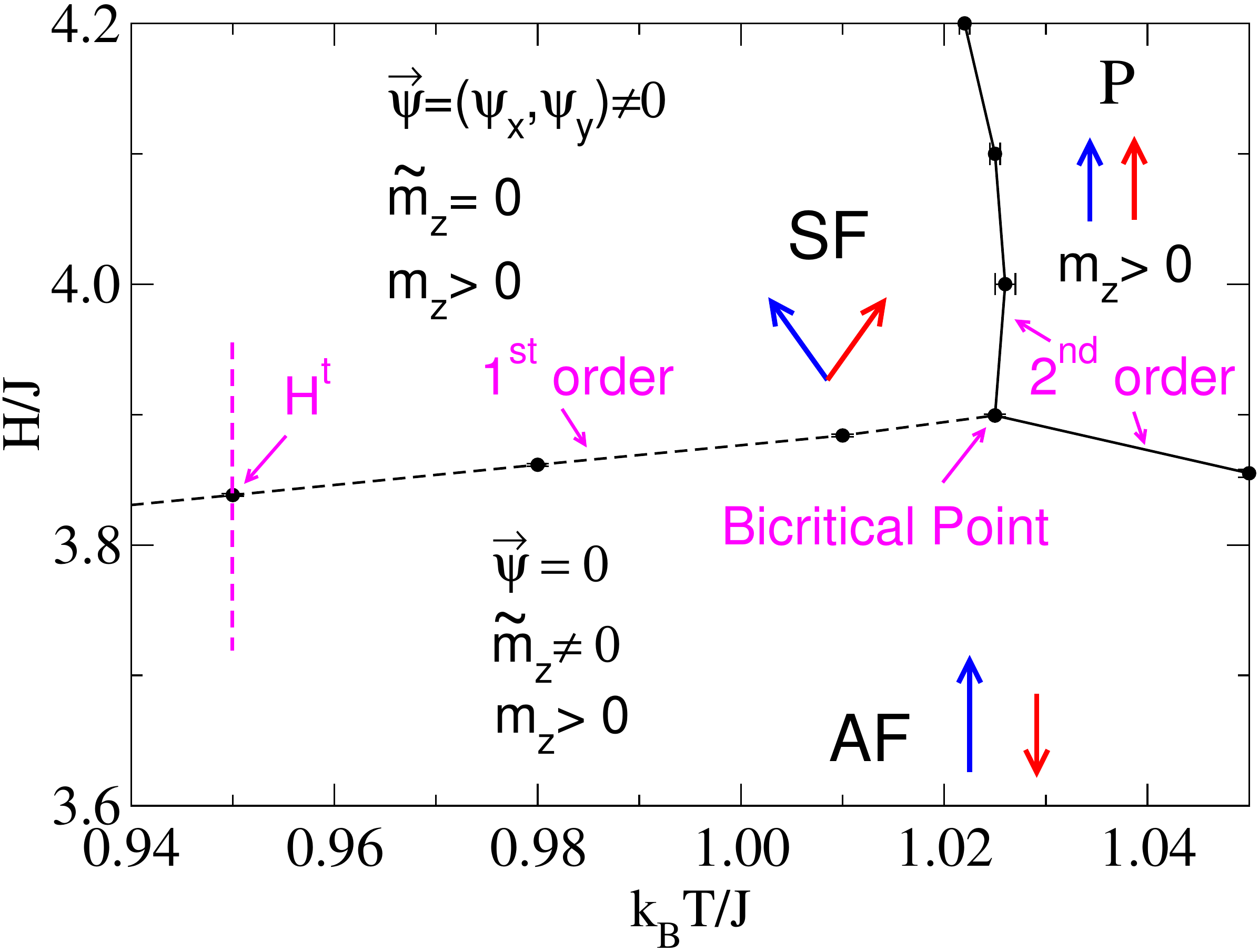}
\caption{\label{phasdiag}  Phase diagram for the simple cubic, anisotropic Heisenberg model in an applied magnetic field, $H$, near the bicritical point~\cite{HTL}.  Both the field and the temperature, $T$, ($k_B$ is Boltzmann's constant) are normalized by the exchange constant $J$.  The order parameter for the antiferromagnetic phase is $\tilde{m_z}$ and for the spin-flop phase is $\vec\psi$.  The z-component of the uniform magnetization is $m_z$.}
\end{figure}

The SF to P and AF to P phase transition lines are of second order and belong to the XY and the Ising universality classes, respectively.  In contrast, a  line of first-order transitions separates the AF and SF phases. The point $T =T_b$ where the three phases meet was determined to be a bicritical point in the three-dimensional (3D) Heisenberg universality class. The resultant phase diagram in the vicinity of the bicritical point is shown in Fig. \ref{phasdiag}.  In earlier work the spin-flop boundary for the anisotropic Heisenberg antiferromagnet in a field was located rather precisely ~\cite{WS,HTL}, so this model is indeed a fertile testing ground for the study of finite size effects at a first order transition where a continuous symmetry is broken. 

 In the next section we will clearly define the model to be studied, describe the Monte Carlo methodology used to generate data for different lattice sizes, and provide a theoretical formulation for the understanding of the finite size behavior of this model (mathematical details will be assigned to Appendix A).  In Sec.III we will present the results of careful Monte Carlo simulations which are then used to test the theory.   Our conclusions are summarized in Sec.IV.

\section{Model and Methods}
\subsection{Model}
The XXZ antiferromagnetic model studied here is described by the Hamiltonian in Eq. (\ref{hamil}). 

\begin{equation} \label{hamil}
\mathcal{H} =  J \sum\limits_{\langle i, j\rangle} [\Delta (S_{ix} S_{jx} + S_{iy}S_{jy}) + S_{iz} S_{jz}] - H \sum\limits_i S_{iz}   
\end{equation}

\noindent where the classical spins {\bf S}$_i$ are unit vectors with components $(S_{ix},S_{iy},S_{iz})$ on sites $i$ of a simple cubic lattice with linear size $L$, and $J>0$ is the exchange coupling between nearest-neighbor pairs of spins. The first summation is over all $\langle i,j\rangle$ pairs of nearest-neighbor sites and the second summation is over all $N=L^3$ spins on the lattice. $\Delta$ is the uniaxial 
exchange anisotropy, which we set to $\Delta = 0.8$ in this work. An external magnetic field $H$ is applied along the $z-$axis, which is the easy axis of the model. 

The phase diagram in the temperature $T$ and field $H$ plane is shown in Fig. \ref{phasdiag}, where the meeting point of the three phase transition lines is estimated to be \cite{WS,HTL} at  $k_BT_b/J = 1.025 \pm 0.0025$ and $H_b/J = 3.89 \pm 0.01$.   Note that here we are not concerned with the phase boundaries near the bicritical point but rather shall study the finite size effects associated with the first order transition from the AF phase to the SF phase at $T=0.95J/k_B$ at a transition field $H^t$.

\subsection{Monte Carlo Methods}
To carry out Monte Carlo simulations we used two different Monte Carlo methods  and considered simple cubic lattices with even values of $L$  and periodic boundary conditions.  Simulations for $L \leq 60$ were first performed
using a standard Metropolis algorithm \cite{metropolis} with the R1279 shift register random number generator~\cite{MCbook}.  Runs of length $3\times10^7$ MCS were performed for all lattice sizes and the number of independent runs ranges from $10$ for $L = 30$ to  $1035$ for $L  = 60$.  
For $L = 60$ Metropolis sampling had great difficulty tunneling between the two states on opposite sides of the spin-flop transitions. Therefore, to insure that the sampling was truly ergodic for $L=60$, $L=80$ and $L=100$, we implemented multicanonical sampling~\cite{BergNeuhaus}.  Multicanonical simulations were performed for the entire range of sizes so that results could be compared with those from Metropolis sampling. 
The multicanonical sampling probability was determined iteratively for each $L$ and then runs of length $10^7$ MCS were carried out.  To determine averages and error bars a total of  $100$ independent runs were made for $L = 30$ and the number increased with increasing size until $900$ independent runs were used for $L =100$.  For smaller lattices, there was agreement between the data generated using the two different sampling methods and the results could be combined for the analysis.
For the multicanonical runs the Mersenne Twister random number generator was used~\cite{MCbook}.   

We employed histogram reweighting techniques \cite{reweighting} to extract thermodynamic quantities for fields near the values used in the simulations.  For the largest lattices, comparisons were made between runs made at adjacent field values and reweighted results to insure that we were not reweighting beyond the reliable range of fields.

As we shall see shortly only data for $L \ge 40$ were in the asymptotic finite size scaling regime, so smaller lattice data were omitted from plots of raw data that will be shown in the results section of this manuscript.

\subsection{Theory}
\subsubsection{General relations}
For small enough fields, $H < H^t (T)$, and low temperatures, $T$, the anisotropic Heisenberg antiferromagnet exhibits Ne\'el-type two-sublattice order on the simple cubic (or other bipartite three-dimensional) lattices. This order is described by the staggered magnetization (with the two interpenetrating  sublattices of the $L \times L \times L$ lattice denoted by indices $1$ and $2$ )

\begin{equation} \label{eq2}
\tilde{m}_ z= \frac{1}{L^3} \Big(\sum\limits_{i \in 1} S_{iz} - \sum\limits_{i \in 2} S_{iz} \Big) \enskip.
\end{equation}

\noindent For $H > 0$ we also expect to have a non-zero uniform magnetization $m_z$ (per spin)

\begin{equation} \label{eq3}
m_z=\frac{1}{L^3} \Big(\sum\limits_{i \in 1} S_{i z} + \sum\limits_{i \in 2} S_{iz}\Big) \enskip.
\end{equation}

\noindent At the transition field, $H=H^t(T)$, there is a first-order phase transition to the ``spin-flop'' phase described by a two-component order involving the transverse spin components

\begin{equation} \label{eq4}
\psi _\alpha =\frac{1}{L^3} \Big(\sum\limits_{i \in 1} S_{i \alpha} - \sum\limits_{i \in 2} S_{i \alpha} \Big) \enskip , \quad \alpha = (x,y) \enskip .
\end{equation}

\begin{figure}
\centering
\includegraphics[angle = 0, width=0.65\hsize]{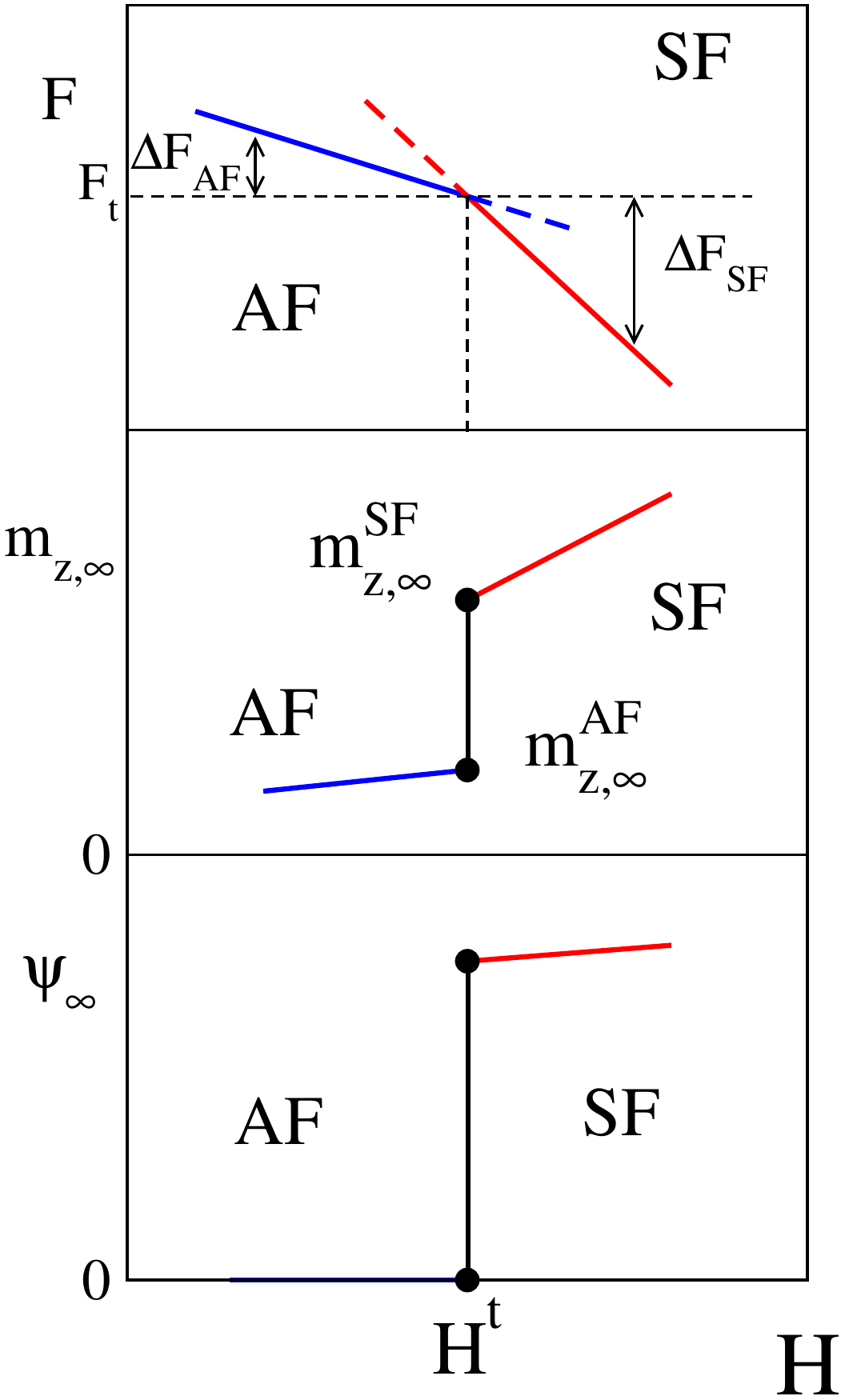}
\caption{\label{fig2}
Schematic variation of thermodynamic quantities with field $H$ for  $T < T_b$:  (top) Free energy $F(T, H)$. The absolute magnitude of the free energy differences $\Delta F_{AF}$, $\Delta F_{SF}$ in the antiferromagnetic (AF) and spin-flop (SF) phases are relative to the free energy (per spin) at the transition $(F_t)$.  Near the transition field $H^t$ these free energy differences vary linearly with $H$; (middle) Magnetization $m_{z, \infty}$ along the field direction.  At $H^t$ a jump occurs from $m^{AF}_{z, \infty}$ (in the thermodynamic limit, $L \rightarrow \infty$) in the AF phase to $m^{SF}_{z, \infty}$ in the SF phase. Near $H^t$ the variation of $m_{z,\infty}$ with $H$ is linear; (bottom) Absolute value of the order parameter of the SF phase, $\psi_\infty$.  Since the AF-SF transition is first order, $\psi_\infty$ jumps discontinuously from zero to a non-zero value when $H = H^t$. }
\end{figure}

\noindent Both $\psi_x$, $\psi_y$ are equivalent and form the components of a vector order parameter $\vec{\psi}$ with XY symmetry.  Fig.~\ref{phasdiag} shows the phase diagram, and Fig.~\ref{fig2} describes the schematic variation of the free energy with the magnetic field $H$ (for simplicity, the temperature dependence of $H^t(T)$ is suppressed in Fig.~\ref{fig2}) along with its derivative $m_{z, \infty} =\langle m_z\rangle _{T, L \rightarrow \infty}$ as well as the order parameter $\psi_\infty=\sqrt{\langle \psi^2_x + \psi ^2_y \rangle_{T, L \rightarrow \infty}}$.  Henceforth, a subscript ``$\infty$'' means that we are referring to properties in the thermodynamic limit.
Note that the variable $m_{z, \infty}$ is the thermodynamically conjugate variable to the magnetic field, with $F$ being the Gibbs free energy per spin~\cite{free_energy},

\begin{equation} \label{eq5}
m_{z, \infty} =- (\partial F/\partial H)_T \enskip .
\end{equation}

Defining the transition field in the thermodynamic limit as $H^t$, we note that $m_{z, \infty}$ at $H^t$ must jump from $m^{AF}_{z, \infty}$ in the antiferromagnetic phase to a larger value $m^{SF}_{z, \infty}$ in the spin-flop phase. At the transition field the free energies of the two phases are equal, $F=F_t$, and  the free energy differences (per spin) in the two phases relative to this value are, to leading order

\begin{equation} \label{eq6}
\Delta F_{AF} =(H^t - H)m^{AF}_{z, \infty} \enskip ,\quad H \leq H^t \enskip,
\end{equation}

\begin{equation} \label{eq7}
\Delta F_{SF} =(H^t - H)m^{SF}_{z, \infty} \enskip ,\quad H \geq H^t \enskip.
\end{equation}

\noindent The total free energy difference between phases then becomes (continuing both phases into their metastable region and ignoring the inequalities in Eqs.~(\ref{eq6}),~(\ref{eq7}))

\begin{eqnarray} \label{eq8}
&&\Delta F \equiv \Delta F_{SF} -\Delta F_{AF} =(H - H^t) (m^{AF}_{z, \infty} - m^{SF}_{z, \infty})\nonumber \\
&&\qquad\qquad\qquad\qquad\qquad =-(H-H^t)\Delta{m}
\end{eqnarray}

\noindent where we have introduced $\Delta{m}=(m^{SF}_{z, \infty} - m^{AF}_{z, \infty})$ to represent the jump in the magnetization $m_z$ in the thermodynamic limit at the transition (see Fig.~\ref{fig2}).  Note that $\Delta F > 0$ for $H < H^t$ since $m^{AF}_{z, \infty} < m^{SF}_{z, \infty}$.

We can use Eqs.~(\ref{eq6}) - (\ref{eq8}) to construct the statistical weights of the two phases in a large, but finite, system in a field $H \approx H^t$. The most naive assumption would be $a_{AF} \propto \exp (-\Delta F_{AF} L^3/k_BT)$, $a_{SF} \propto \exp (-\Delta F_{SF} L^3/k_BT)$.  However, this assumption, using a common normalization factor, disregards the difference in degeneracy of the two phases.  While the $A F$ phase is two-fold degenerate ($n=1$, one-component order parameter), in the SF phase a continuous (XY-model like) symmetry is broken ($n=2$, two-component order parameter).  How this difference enters in the weights is not obvious, unlike in the simpler case of the thermally driven $q$-state Potts model where the high-temperature phase is non-degenerate and the low-temperature phase is simply $q$-fold degenerate. There, an extra factor $q$ appears in the weight of the low-temperature phase multiplying the Boltzmann factor.

It  is unclear (at least to us) what this factor $q$ must be when dealing with a continuous symmetry. Thus, we introduce an analogous factor $q$ $\it phenomenologically$ to account for the difference in degeneracy between the phases at the coexistence point in the thermodynamic limit $(H=H^t)$. As we shall see, in full analogy with the Potts model, this factor leads to shifts of characteristic finite size induced features (e.g. position of the maximum of the magnetic susceptibility or specific heat, minimum of the various cumulants, etc.). Observing these features in the simulations should provide numerical estimates for this degeneracy factor $q$ for the present problem.  In the next subsection we shall advance a hypothesis for the value of $q$.

We now postulate the ``equal weight rule''~\cite{BK} for the statistical weights of the two phases, i.e. $a_{AF} = \mathcal{N} \exp (-\Delta F_{AF} L^3/k_BT)$ and $a_{SF} =\mathcal{N}q \exp (-\Delta F_{SF} L^3/k_BT)$, with a normalization factor $\mathcal{N}$.  Requiring $a_{AF}+ a_{SF}=1$ yields

\begin{equation} \label{eq9}
a_{AF} = \exp (\Delta F L^3/k_BT)/[q +\exp (\Delta F L^3 / k_BT)] \enskip,
\end{equation}

\begin{equation} \label{eq10}
a_{SF} = q/[q + \exp (\Delta FL^3 /k_BT)] \enskip ,
\end{equation}

\noindent as expected.

Formally, the factor $q$ can be absorbed by redefining the weights as $a_{AF} =\exp (\Delta F' L^3/k_BT)/[1 + \exp (\Delta F' L^3/k_BT)]$, $a_{SF} =1/[1 +\exp (\Delta F' L^3/k_BT)]$, with $\Delta F'=\Delta F- (k_BT/L^3)
\ln q$. This shows that finite size induced shifts of characteristic features scaling as ($k_BT/L^3) \ln q$ will occur.

\subsubsection{Order parameter distribution at the transition in the thermodynamic limit}
Before going further, we use the transition between the disordered and ordered states for the $q$-state Potts model to provide some insight about the effective degeneracy factor for the AF to SF transition in the anisotropic Heisenberg model. For the $q$-state Potts model in the thermodynamic limit exactly at the transition temperature, the probability distribution of the order parameter $P(\vec\psi)$ is simply the sum of $q+1$ weighted delta functions,

\begin{equation}\label{Ppsi}
P(\vec\psi)=\delta(\vec\psi)+\sum_{k=1}^{q}\delta(\vec\psi-\vec\psi_k)
\end{equation}

\noindent where the $\vec\psi_k$ are the discrete values of the order parameter in the ordered phase~\cite{Wu}.  The first term on the right hand side of  Eq.~(\ref{Ppsi}) represents the disordered phase and the second term represents the (degenerate) ordered phase.
In the current case a similar expression holds except that the $\vec\psi$ are continuous.  We, therefore, conjecture that

\begin{widetext}
\begin{equation}\label{coex}
P(\tilde{m}_z,\psi)=[\delta(\tilde{m}_z-\tilde{m}_{z,\infty})+\delta(\tilde{m}_z+\tilde{m}_{z,\infty})]\delta(\psi) +\int_{0}^{2\pi}\delta(\tilde{m}_z)\delta(\psi-\psi_\infty)d\phi
\end{equation}
\end{widetext}

\noindent where the order parameter $\vec\psi$ is written in terms of the magnitude $\psi$ and angle $\phi$ in the $(\psi_x,\psi_y)$ plane and we have integrated over $\phi$.  The index ``$\infty$'' indicates that the thermodynamic limit was taken first and then $H \to H^t$. 
Since there is no dependence upon $\phi$, the integral gives $2\pi$.  Integration over $\tilde{m}_z$ then yields

\begin{equation}\label{coex'}
P(\psi)=2\delta(\psi)+2\pi\delta(\psi-\psi_\infty) \enskip,
\end{equation}

\noindent or the relative weight of the two phases is simply $\pi$!

Eq.~(\ref{coex}) merely indicates that in the thermodynamic limit and for $H=H^t$, we have
phase coexistence between pure AF phases ($\tilde{m}_z= \pm \tilde{m}_{z,\infty}$, $\vec{\psi}=0$) and pure SF phases ($\vec{\psi}=(\psi_\infty, \phi)$, in polar coordinates in the ($\psi_x, \psi_y)$-plane, and $\tilde{m}_z=0$). The distribution of the order parameters is simply characterized by the appropriate Dirac delta-functions. Making contact with formulation of Eqs.~(\ref{eq9}),~(\ref{eq10}), where the relative weights of the two phases at $H^t$ was denoted by the phenomenological parameter $q$, we find that the joint (unnormalized) distribution of the order parameters $\tilde{m}_z$, $\psi=|\vec{\psi}|$ becomes \\

\begin{widetext}
\begin{equation} \label{eq32'}
P_\infty (\tilde{m}_z, \psi)= [\delta (\tilde{m}_z-\tilde{m}_{z,\infty}) + \delta (\tilde{m}_z + \tilde{m}_{z, \infty})] \delta (\psi) + 2 q \delta (\tilde{m}_z) \delta (\psi- \psi_\infty) \enskip .
\end{equation}\\
\end{widetext}

\noindent The normalization constant for this distribution is

\begin{equation} \label{eq32''}
\mathcal{N}_\infty =\int\limits^{+1}_{-1} d \tilde{m}_z \int\limits_{0^-}^1 d \psi P_\infty (\tilde{m}_z, \psi)=2+2q \enskip .
\end{equation}

From Eqs.~(\ref{eq32'}),~(\ref{eq32''}) we can easily obtain the moments and cumulants of both order parameters (the notation $\langle \cdots \rangle_\infty$ means that an average over both phases at the transition point in the thermodynamic limit is taken)

\begin{eqnarray} \label{eq32'''}
&&\langle |\vec\psi| \rangle_\infty= \psi_{\infty} q /(1+q) \enskip,\\
&&\langle \psi^2 \rangle_\infty = \psi^2_\infty q/(1+q) \enskip, \\
&&\langle \psi^4 \rangle_\infty  = \psi^4_\infty q/(1+q) \enskip ,
\end{eqnarray}

\begin{eqnarray} \label{eq32IV}
&&\langle |\tilde{m}_z| \rangle_\infty =\tilde{m}_{z, \infty} /(1+q) \enskip,\\
&&\langle \tilde{m}^2_z \rangle_\infty =\tilde{m}^2_{z, \infty} /(1+q) \enskip,\\ 
&&\langle \tilde{m}_z^4 \rangle _\infty =\tilde{m}^4_{z, \infty}/(1+q) \enskip .
\end{eqnarray}

Hence, the cumulants simply become

\begin{equation} \label{eq32V}
U^{xy}_\infty \equiv 1- \langle \psi^4 \rangle_\infty /[3 \langle \psi^2 \rangle^2_\infty ]=1 - \frac{1+q}{3q} \enskip,
\end{equation}

\noindent and

\begin{equation} \label{eq32VI}
U_\infty ^z \equiv 1 - \langle \tilde{m}^4_z \rangle_\infty /[3 \langle \tilde{m}^2_z \rangle^2 _\infty] =(2-q)/3 \enskip .
\end{equation}

Of course, these results do not invoke the assumption of Gaussian distributions of the order parameters for finite $L$ \{see Eqs.~(\ref{eq11}),~(\ref{eq14}),~(\ref{eq19}) or a similar assumption for $P_L (\tilde{m}_z)$ that will be used below\}. Thus, Eqs.~(\ref{eq32'''})-(\ref{eq32VI}) are not affected in any way by deviations from Gaussian distributions in the wings of the actual distributions for finite $L$.

Eqs.~(\ref{eq32'''})-(\ref{eq32VI}) permit stringent tests of this theory by simulations using the following recipe:  Suppose an accurate estimate of $H^t$ is known from suitable finite size extrapolation (e.g. using Eq.~(\ref{eq13a}); other choices will be given below). Then, a very large system can be simulated (for which no transitions between the pure phases occur for very long runs) right at $H=H^t$, once starting in the AF phase and once starting in the SF phase, to obtain very accurate estimates of $\tilde{m}_{z, \infty}$ and of $\psi_\infty$.  From the distributions of the order parameters in these pure phases, accurate estimates of the staggered susceptibilities $\tilde{\chi}^{AF}_{xy}$, $\tilde{\chi}^{SF}_{xy}$, $\tilde{\chi}^{AF}_{zz}$, and $\tilde{\chi}^{SF}_{zz}$, can also be extracted. Extrapolations of the estimates for
$\langle |\vec\psi| \rangle_L$, $\langle |\tilde{m_z}| \rangle_L$,
 $\langle \psi^2 \rangle_L$, $\langle \tilde{m}^2_z \rangle_L$, $U^{xy}_L$ and $U^z_L$ at $H=H^t$ towards $L=\infty$ should provide estimates for the factors $q/(1+q)$ and $1/(1+q)$ in Eqs.~(\ref{eq32'''}),~(\ref{eq32IV}) as well as the cumulants, Eqs.~(\ref{eq32V}),~(\ref{eq32VI}).
 
Eqs. (\ref{eq32'''}) -~(\ref{eq32VI}) also define (almost) universal intersection points when we study analogous averages for finite $L$ as a function of the field $H$.  $\langle |\vec\psi| \rangle_L$, $\langle | \tilde{m}_z | \rangle_L$,
 $\langle \psi^2 \rangle_L$, $\langle \tilde{m}^2_z \rangle_L$, $U^{xy}_L$ and $U^z_L$  are all analytic functions of $H$, saturating for $|\Delta F|L^3/k_BT >>1$.  For the cumulants these saturation values are trivial, e.g. $U^z (H<H^t)=1$, $U^z (H>H^t)=0$.  But we will show later that they will agree with Eqs. (\ref{eq32'''}) -~(\ref{eq32VI}), up to corrections of order $L^{-\frac{3}{2}}$ or $L^{-3}$ for $|\Delta F|L^3/k_BT =0$.  Unlike second-order transitions for which only cumulants, e.g. $U_L^{xy}$ and $U_L^z$, have unique intersection points at the transition, for this first order transition both the individual moments and the cumulants exhibit this feature of common intersection points at $H=H^t$.

\subsubsection{Two-Gaussian approximation for the magnetization distribution}
While Eqs.~(\ref{eq32'''})-(\ref{eq32VI}) describe the behavior of the system when we first set $H=H^t$ and then take the limit $L \rightarrow \infty$, it is also of great interest to explore the leading corrections to the limiting behavior when $L$ is large but finite.

Following general considerations of statistical physics~\cite{L&L}, in pure phases, for large but finite size systems, we expect Gaussian distributions for the densities of extensive thermodynamic variables rather than $\delta$-functions.  A simple case to consider is the uniform magnetization for which Gaussian distributions for the (scalar) quantity $m_z$ in the two phases would give a distribution,

\begin{eqnarray} \label{eq11}
&& P_L(m_z) \propto \frac{a_{AF}}{\sqrt{\chi^{AF}_{zz}}} \exp \Big\{- \frac{[m_z-(m^{AF}_{z,{\infty}} + \chi^{AF}_{zz} \Delta H)]^2 }{2 k_BT \chi^{AF}_{zz} /L^3} \Big\}\nonumber\\
&&\quad+ \frac{a_{SF}}{\sqrt{\chi^{SF}_{zz}}} \exp \Big\{-\frac{[m_z - (m^{SF}_{z,{\infty}} + \chi^{SF}_{zz} \Delta H)]^2}{2 k_BT \chi^{SF}_{zz}/ L^3}\Big\} \enskip,
\end{eqnarray}

\noindent where

\begin{equation} \label{eq12}
\Delta H \equiv H-H^t
\end{equation}

\noindent and $\chi^{AF}_{zz}, $ $\chi^{SF}_{zz}$ are the susceptibilities at $(H=H^t)$ in the two phases. 

Invoking the analogy of these equations to the case of the Potts model energy distribution [cf. Eq.~(IV.21) in Ref.~\cite{MCbook}], we conclude that the susceptibility peak should scale as

\begin{equation} \label{eq13}
\chi^{\rm max}_{zz} \approx \frac{\chi^{AF}_{zz} +\chi^{SF}_{zz}}{2} + \frac{(\Delta{m})^2 L^3}{4 k_BT} 
\end{equation}

\noindent in analogy to the specific heat of the Potts model.  Note that the location of this maximal magnetization fluctuation (i.e. susceptibility of the z-component of the uniform magnetization) occurs when the two weights are equal, i.e.~$a_{AF} =a_{SF} =1/2$. This condition readily yields $q \exp (-\Delta F L^3/k_BT)=1$, i.e.~$\Delta F/k_BT=\ln q/L^3$, or 

\begin{equation}\label{eq13a}
(H^{\rm max} - H^t)/k_BT=-[\Delta{m}L^3]^{-1} \ln q \enskip. 
\end{equation}

\noindent Since $m^{AF}_{z, \infty} < m ^{SF}_{z, \infty}$, the position of the susceptibility maximum relative to the transition point must shift to smaller fields, $H^{\rm max} < H^t$, and scale with size like $L^{-3}$. 

The susceptibility at the transition point $H^t$  is smaller by a factor $4q/(1+q)^2$ than $\chi_{zz}^{\max}$ for $L \to \infty$.

One important caveat, however, is that for the weight $a_-$ for the low temperature phase of the Potts model, the factor $q$ reflecting the degeneracy of the ordered phase is known. In contrast, here the value of a similar factor representing the difference in degeneracies of the spin-flop and antiferromagnetic phases is $\it unknown$ unless we rely on the hypothesis of Eq.~(\ref{coex'}) that $q=\pi$.

\subsubsection{The SF phase order parameter distribution}
We next consider the distribution of the SF order parameter $\vec{\psi}$.
For $H< H^t$, i.e. in the AF phase, there is simply a Gaussian distribution about zero since the transverse spin component is disordered,

\begin{equation} \label{eq14}
P^{AF}_L (\vec{\psi}) = \mathcal {N} \exp \Big(-\frac{{\vec\psi} {{\:}^2}}{2 k_BT \tilde{\chi}^{AF}_{xy} /L^3} \Big) \enskip,
\end{equation}

\noindent where we introduced the notation $\tilde{\chi}_{\alpha \beta}$ for the tensor of staggered susceptibilities, and $\tilde{\chi}^{AF}_{xy}$ stands for the $xy$-components of the staggered susceptibility in the AF phase.

The order parameter distribution in the SF phase is more interesting, 

\begin{equation} \label{eq19}
P^{SF}_L (\vec{\psi})= \mathcal{N} \exp \Big[-\frac{({\vec\psi} {{\:}^2} - \psi^2_\infty)^2}{8{\psi^2_\infty} k_BT \tilde{\chi}^{SF}_{xy}/L^3}\Big] \enskip,
\end{equation}

\noindent where now $\tilde{\chi}^{SF}_{xy}$ denotes the $xy$-component of the staggered susceptibility in the spin-flop phase.  Note that a 4th order polynomial in $\vec\psi$ is needed in the argument of the exponential function in Eq.~(\ref{eq19}) to bring out the spherical symmetry in the $(\psi_x,\psi_y)$-plane correctly.  Near the peak $(|\vec\psi|\approx \psi_{\infty})$ the argument of the exponential reduces to the expected quadratic form, i.e. $-(\psi-\psi_{\infty})^2/(2k_BT\tilde\chi_{xy}^{SF}/L^3)$.

In the vicinity of the transition field, $H^t$, we now make the standard superposition approximation,

\begin{eqnarray}\label{eq32a}
&&\langle\psi^2\rangle_L=a_{AF}\langle\psi^2\rangle_{AF}+(1-a_{AF})\langle\psi^2\rangle_{SF}\\
\label{eq32b}
&&\langle\psi^4\rangle_L=a_{AF}\langle\psi^4\rangle_{AF}+(1-a_{AF})\langle\psi^4\rangle_{SF}
\end{eqnarray}

\noindent where

\begin{equation}\label{eq32a'}
a_{AF}=1/[1+q\exp(\mathcal{Z})] \enskip.
\end{equation}

\noindent We have written $\mathcal{Z}$ =$-\Delta{F}L^3/k_BT$, and the moments $\langle \cdots \rangle_{AF},$ $ \langle \cdots \rangle _{SF}$ refer to the order parameter distributions in the ``pure'' AF and SF phases respectively.

Of course, we could repeat the calculation of Eqs.~(\ref{eq32'})-(\ref{eq32VI}) for finite $L$, replacing the delta functions by the appropriate Gaussian distributions, e.g. Eqs.~(\ref{eq14}) and ~(\ref{eq19}). These calculations are straightforward, but clumsy, so we confine the details to Appendix A and only give a few final results here.

From the general result for the fourth order cumulant (see Appendix A) we immediately conclude that for $H^t$, $U^{xy}_L$ differs from $U^*$=$U^{xy}_\infty$ only by corrections of order $L^{-3}$ and

\begin{equation}\label{eq33aa}
U^{xy}_L \mid_{H^t} =1 - \frac{1 + q} {3q} \Big[1+( \frac{4k_B T \tilde{\chi}^{SF}_{xy}} {\psi^2_\infty} - \frac{4 k_BT \tilde{\chi}^{AF}_{xy}}{q \psi^2_\infty})  \frac{1}{L^3}\Big] \enskip .
\end{equation}
\\

\noindent If we use the value $q=\pi$, we find that $U_\infty^{xy}\approx0.56056$.

Taking the derivative of the expression for the cumulant to find the minimum, and writing $Y$=$q\exp(\mathcal{Z})$, we find that the location of the minimum is given by $Y_{\min}\approx 2k_BT\tilde\chi^{AF}_{xy}/((\psi_{\infty})^2L^3)$.  Hence

\begin{equation}
H_{\min}=H^t-\frac{k_BT[{\ln}q-{\ln}Y_{\min}]}{\Delta{m}L^3} \enskip.
\end{equation}

\noindent Thus, the shift in the location of the minimum scales as $L^{-3}$,  but the leading term is actually ${\ln}L/L^3$.  At the minimum, the value of $U_L^{xy}$ is

\begin{equation}
U_{L,\min}^{xy}\approx{\it const}-\frac{\psi_\infty^2L^3}{24k_BT\tilde\chi_{xy}^{AF}}
\end{equation}

\noindent which means that $U_{L,\min}^{xy}$ approaches $-\infty$ proportional to $-L^3$ as $L$ approaches infinity.\\

\subsubsection{AF phase order parameter distribution}
The AF order parameter distribution for this one-component order parameter in the AF phase is a double Gaussian, analogous to Eq.~(\ref{eq11}),

\begin{eqnarray}\label{eq35'}
&&P^{AF}_L(\tilde{m_z})\propto  \exp \Big[{- (\tilde{m_z}-\tilde{m}_{z,{\infty}})^2L^3/(2k_BT \tilde\chi^{AF}_{zz}) }\Big]\nonumber\\
&&\qquad +\exp \Big[{- (\tilde{m_z} +\tilde{m}_{z,{\infty}})^2L^3/(2k_BT \tilde\chi^{AF}_{zz}) }\Big]
\end{eqnarray}

\noindent whereas in the SF phase it is given by a single Gaussian 

\begin{equation}\label{eq35''}
P^{SF}_L(\tilde{m_z})= {\sqrt{L^3/{(2{\pi}k_BT \tilde{\chi}^{SF}_{zz}}})} \exp \Big[-\tilde{m}_z^2L^3/(2k_BT \tilde{\chi}_{zz}^{SF})\Big]
\end{equation}
\\

\noindent From these distribution and superposition approximations analogous to Eqs.~(\ref{eq32a}) -~(\ref{eq32a'}), it is straightforward to calculate the various moments of the staggered magnetization and the cumulant (see Appendix A).  Then, at the transition, the fixed point value of the cumulant is

\begin{equation}\label{eq33b}
U^{z}_L \mid_{H^t} =\frac{2-q}{3} + \frac{1 + q}{3} \Big[\frac{2q k_BT \tilde{\chi}^{SF}_{zz}}{\tilde{m}^2_{z, \infty}} - 4 \frac{k_BT \tilde{\chi}^{AF}_{zz}} {\tilde{m}^2 _{z, \infty}} \Big]\frac{1}{L^3} \enskip.
\end{equation} 
\\

\noindent This means that, as expected, the asymptotic value of the cumulant coincides with the result from the treatment using delta function distributions.  Again the correction to this result is small, of order $L^{-3}$; and choosing $q=\pi$, we find the fixed point value of the cumulant is $U_*^{z} \approx -0.38$.\\

%



\subsubsection{Maximum slope of the cumulant and cumulant crossings}

We have shown that the cumulant at $H=H^t$ is of order unity but reaches a deep minimum (of order $-L^3$ for $L \rightarrow \infty$) at a value $H^t- H \propto L^{-3}$ (on the AF side of the spin-flop transition when the cumulant of the SF-order parameter is considered).  In order to achieve a variation of order $L^{+3}$ in an interval of order $L^{-3}$, the maximum slope of the cumulant in the interval between the minimum and the crossing point must then be of order $L^6$.
On the other hand, we can easily show that the slope of the cumulant right at $H=H^t$ still is only of order $L^3$, namely

\begin{equation}\label{eq99}
k_BT \,{\frac{d U^{xy}_L}{d H}} \biggr\vert_{H^t} =\frac{1}{3q} \Delta{m} L^3 \enskip.
\end{equation}

\noindent Thus, significant curvature should appear in the plot of $U_L^{xy}$ vs. $H$ near $H^t$, since the slope first increases from zero at the cumulant minimum to a value of order $L^6$ and then decreases to a value of order $L^3$ at $H=H^t$. The location of the intersection point, which corresponds to a cumulant value of order unity, hence must be very close to $H^t$ ($|H^{\rm cross}-H^t|\propto L^{-6}$) also. On a scale of $H-H^t$ of order $L^{-3}$ (the regime over which the variable $\mathcal{Z}$ exhibits a significant variation) differences in cumulant intersections of order $L^{-6}$ are completely negligible.  Consequently, the spread of the cumulant intersections is very small, although in a strict sense there is no unique cumulant crossing point. Two cumulants for linear dimensions $L$ and $L' =L + \delta$ that differ slightly can be shown to cross for $H^{\rm cross}$ given by

\begin{widetext}
\begin{equation}\label{eq111}
\frac{H^{\rm cross} - H^t}{k_BT} \Delta{m}=\frac{1+q}{q} \Big[4 k_BT \tilde\chi^{AF}_{xy}/\psi^2_\infty - 4qk_BT \tilde\chi^{SF}_{xy}/\psi^2_{\infty} \Big] L^{-6} \enskip.
\end{equation}
\end{widetext}

\noindent This equation is derived by a Taylor expansion of $U_{L + \delta}$ simultaneously in the small variables $H-H^t$ and $\delta/L$. This result verifies the above argument that the scale for the shift of the cumulant crossing is negligibly small in comparison with that for the shift of the cumulant minimum.  The shift of the intersection point thus scales as the square of the inverse volume, while the regime over which the transition is spread out is given by

 \begin{equation} \label{eq41}
 \Delta \mathcal{Z} =1 \enskip, \quad \Delta H/k_BT=1 / [\Delta{m}L^3 ] \enskip .
 \end{equation}

\noindent The situation is analogous to the case of the (temperature driven) transition in the Potts model, cf. Vollmayr et al. ~\cite{Vollmayr}. As in the latter case, and unlike the simple, field driven first order transition of an Ising ferromagnet below $T_c$, there is no ``equal height rule'' of the order parameter distribution at the transition.  Eqs.~(\ref{eq14}), (\ref{eq15}) show that the peak height at $\vec{\psi}=0$ scales proportional to the volume, $L^3$, while the height of the ``rim'' $|\vec{\psi}|=\psi_{\infty}$ only scales like the square root of the volume, $L^{3/2}$, cf. Eqs. (13), (14), and (23) of Vollmayr at al.~\cite{Vollmayr}

Useful information can be readily extracted about the mean square order parameter $\langle\psi^2\rangle$ as a function of the field $H$ near the transition field $H^t$.

The superposition approximation (Eqs.~(\ref{eq32a}), (\ref{eq32b})) readily shows that $\langle \psi ^2 \rangle$ is described by a simple scaling function of $\mathcal{Z}$, namely

\begin{equation} \label{eq100}
\langle \psi^2 \rangle / \psi^2_\infty=[b + q \exp (\mathcal{Z})] / [1 + q \exp (\mathcal{Z})]
\end{equation}

\noindent where we have introduced the abbreviation

\begin{equation} \label{eq200}
b=(2 k_BT \tilde\chi^{AF}_{xy} ) / (\psi^2_\infty L^3) \enskip .
\end{equation}

Note that the quantities $\psi_\infty$, $\tilde\chi^{AF}_{xy}$ (and $\Delta{m}$ which is needed to convert the scale of $H-H^t$ to $\mathcal{Z}$) can be estimated directly from simulations.  For runs of modest length for very large systems precisely at $H=H^t$, transitions between the phases can be avoided due to metastability.  Starting in the AF and SF states respectively will permit measurements in the pure phases.  ($H^t$ is already known with high precision). For the correct choice of $q$, all choices of (sufficiently large) $L$ should then lead to perfect collapse on a master curve, Eq.~(\ref{eq100}), that is explicitly predicted.

The value of the order parameter at the location of the maximum slope is

\begin{equation} 
\langle\psi^2\rangle=\frac{1}{2}\psi_\infty^2+\frac{{k_B}T\tilde\chi_{xy}^{AF}}{\psi_\infty^2L^3}
\end{equation}

\noindent which is {\bf independent} of the value of ``$q$''.  Since $\psi_\infty$ and $\tilde\chi_{xy}^{AF}$ can be measured independently, we have a non-trivial test of the double Gaussian approximation. The value of the slope at the transition field $H=H^t$ is

\begin{equation} \label{eq3333}
\frac{d\langle\psi^2\rangle}{d(H/k_BT)} \biggr\vert_{H^t} =\Delta{m}L^3\frac{q}{(1+q)^2}(1-\frac{2{k_B}T\tilde\chi_{xy}^{AF}}{\psi_\infty^2L^3})\psi_\infty^2 \enskip.
\end{equation}

We also note that for $H=H^t$, i.e. $\mathcal{Z}=0$, Eq.~(\ref{eq100}) predicts $\langle\psi^2\rangle/{\psi_{\infty}^2}=(b+q)/(1+q) \approx{q/(1+q)}$, as expected from Eq. (17).
  When $\langle\psi^2\rangle/{\psi_{\infty}^2}$ is plotted vs $H$, all curves for large $L$ will intersect for $H=H^t$ at this value.

\subsubsection{Limitations of the phenomenological theory}
At this point, we comment on an important distinction between the order parameter cumulant intersection for first order transitions and for second order transitions.
At second order transitions, for $L \rightarrow \infty$ corrections to scaling can be ignored and the cumulant is a regular function of the variable $\mathcal{Z}=t L^{1/\nu}$, $\nu$ being the correlation length critical exponent. The slope of the cumulant $d U_L/dt$, $t$ being the reduced distance from the transition point, at the transition point hence is of the same order as the inverse shift $L^{-1/\nu}$ of the susceptibility maximum.  

The $L^{-3}$ correction to Eqs.~(\ref{eq33aa}), (\ref{eq33b}) is
simply a ``correction to finite size scaling'' for first order transitions, analogous to those that appear in finite size scaling at second order transitions. However, in the latter case these corrections involve a second, non-trivial exponent (different from $1/\nu$), while here the inverse volume $L^{-3}$ is the only variable that leads to correction terms in the finite size scaling description. Since the resolution of the numerical data is insufficient to quantitatively resolve any of these correction term effects, they shall not be discussed further. We also note that further terms are expected due to corrections to the Gaussian approximation for the probability distribution, e.g.~the result $U^{AF}_L= 1/3$ for the cumulant of the XY-order parameter in the AF phase is also expected to have a $1/L^3$ correction (related to higher order correlation functions) which has been ignored here but seems rather relevant numerically.  In addition, corrections of order $\exp(-L/\xi)$, where $\xi$ is the appropriate correlation length of the AF and/or SF order, have been ignored.  These become important, however, if $H^t$ is close to the bicritical point.

There is another, rather different limitation to our phenomenological treatment:  The superposition approximation, Eqs.~(\ref{eq32a}) and (\ref{eq32b}), which assumes that the total order parameter distribution $P_L(\psi_x, \psi_y$) is a superposition of Gaussians for the AF phase (centered at $\psi_x=0$, $\psi_y=0$) and of the ordered phase (centered at $\psi^2_x + \psi^2_y =\psi^2_\infty$) with appropriate weights, is not accurate away from the peaks of the distribution. The same applies for the distribution of the order parameter $P(\tilde{m}_z)$ for the AF phase. In the latter case, the problem is well understood: far from the transition $(H \ll H^t)$ the distribution $P(\tilde{m}_z)$ near $\tilde{m}_z=0$ is dominated by ``slab configurations'' where a domain with $\tilde{m}_z=-\tilde{m}_z^{\rm spont}$ is separated by two domain walls (``antiphase domain boundaries'') from domains with $\tilde{m}_z= + \tilde{m}_z^{\rm spont}$ (where $\tilde{m}^{\rm spont}_z$ is the value of the order parameter where $P(\tilde{m}_z)$ has its peak).  Thus, $P(\tilde{m}_z)$ is not controlled by a Boltzmann factor containing the volume $L^3$, but rather by the surface area $L^2$,

  \begin{equation} \label{eq46}
  P(\tilde{m}_z \approx 0) \propto \exp (- 2 L ^2 f_{\rm int} /k_BT) \enskip,
  \end{equation}

\noindent with $f_{\rm int}$ the interfacial excess free energy per unit area (interfacial tension). This ``slab configuration'' with two planar interfaces is compatible with the periodic boundary conditions, of course.

For a two-component order parameter, the ``antiphase domain boundaries'' are spread out over the entire volume, and the ``phase'' of the order parameter gradually rotates from zero to $2 \pi$ as the system is traversed. Thus, for $H \gg H^t$, we have

  \begin{equation} \label{eq47}
  P(\psi_x \approx 0, \, \psi_y \approx 0) \propto  \exp (- 2 L \Gamma/k_BT)
  \end{equation}

 \noindent where $\Gamma$ is essentially the ``helicity modulus'' ~\cite{helicity}.  For H near $H^t$, however, mixed phase configurations will occur with states $m_z\approx {m_z^{AF}}$, $\tilde{m}_z\approx\pm{\tilde{m}_{z,
 \infty}}$, $\psi \approx 0$ coexisting with states $m_z\approx {m_z^{SF}}$, $\tilde{m}_z\approx 0$, $|\vec{\psi}| \approx \psi_{\infty}$ with comparable weights.  Such mixed phases require more complex ``interfaces''  in which the order parameters ``interpolate'' between their coexisting phase values.  The generalization of Eqs.~(\ref{eq46}) and~(\ref{eq47}) is unknown.

\section{Monte Carlo Results}
 The simulations were performed below the bicritical point $T_b$ at fixed temperature $T = 0.95J/k_B$, and we varied the external field $H$ in order to determine the phase transition from AF to SF. 
 When not shown, error bars in the figures showing our results are smaller than the size of the symbols.

\begin{figure}
\centering
\includegraphics[clip,angle=270,width=0.95\hsize]{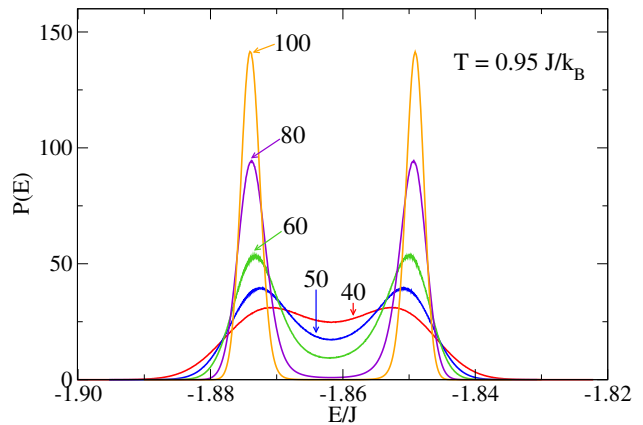}
\caption{\label{probE} Probability distribution (unnormalized) of the energy at the transition field $H^t_L$ for different lattice sizes $L$.}
\end{figure}

The probability distributions of the energy $E$ per site at $T=0.95J/k_B$ are shown at the transition field $H^t_L$ for different lattice sizes $L$ in Fig.~\ref{probE}.  For each size we chose the finite size transition field to be located at the point at which the (symmetric) peaks in the probability distribution for the energy were of equal heights.  While the dip between the peaks was rather shallow for $L=40$, it rapidly became quite deep for increasing values of $L$ in agreement with the predictions of Binder~\cite{Binder1982} and Lee and Kosterlitz~\cite{LeeKosterlitz} for a first order transition.  

For smaller values of $L$ the probability distribution was almost flat, and the resultant thermodynamic properties showed such substantial finite size rounding that it was not possible to extract useful information about the asymptotic behavior.  For this reason, we shall not show raw data for $L <40$ in the figures that follow.

\begin{figure}
\centering
\includegraphics[clip,angle=270,width=0.95\hsize]{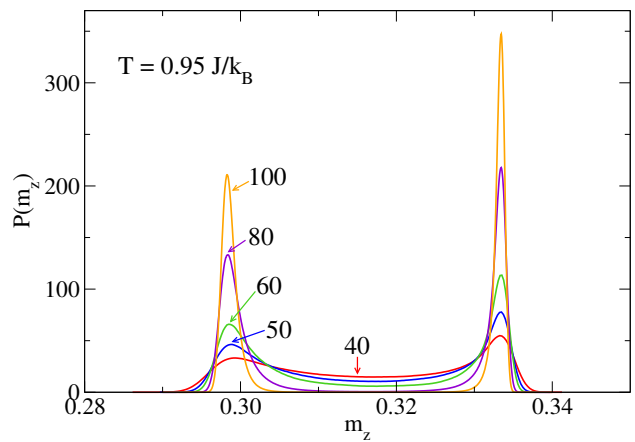}
\caption{\label{probMz} Probability distribution (unnormalized) of the z-component of the magnetization at the transition field $H^t_L$ for different lattice sizes $L$.}
\end{figure}

The probability distributions of the z-component for the magnetization $m_z$ per site at $T=0.95J/k_B$ are shown at the transition field for different lattice sizes $L$ in Fig.~\ref{probMz}.  For each size we chose the finite size transition field to be located at the same point at which the (symmetric) peaks in the probability distribution for the energy were of equal heights.  These data show that the distributions of $m_z$ contain two clear, asymmetric peaks of different heights; moreover, these data cannot be described solely by the sum of two Gaussians. 
Recall that the minimum between the two peaks represents phase coexistence inside the simulation box with one slab in a state $|\tilde{m}_z|\approx {\tilde{m}_{z,{\infty}}}, \psi=0$ and the other slab having $|\tilde{m}_z|\approx 0, |\vec{\psi}|=\psi_{\infty}$ with the two slabs separated by a complex interface connected by the periodic boundary conditions.
 If we ``separate'' the distributions into two peaks by choosing the minimum probability as the separation point, we can measure the ``weight'' of each peak by numerically integrating the probability under each peak.  To a high degree of precision, the peaks for each of the lattices sizes, $L$, then have equal weights.  

\begin{figure}
\centering
\includegraphics[clip,angle=0,width=0.95\hsize]{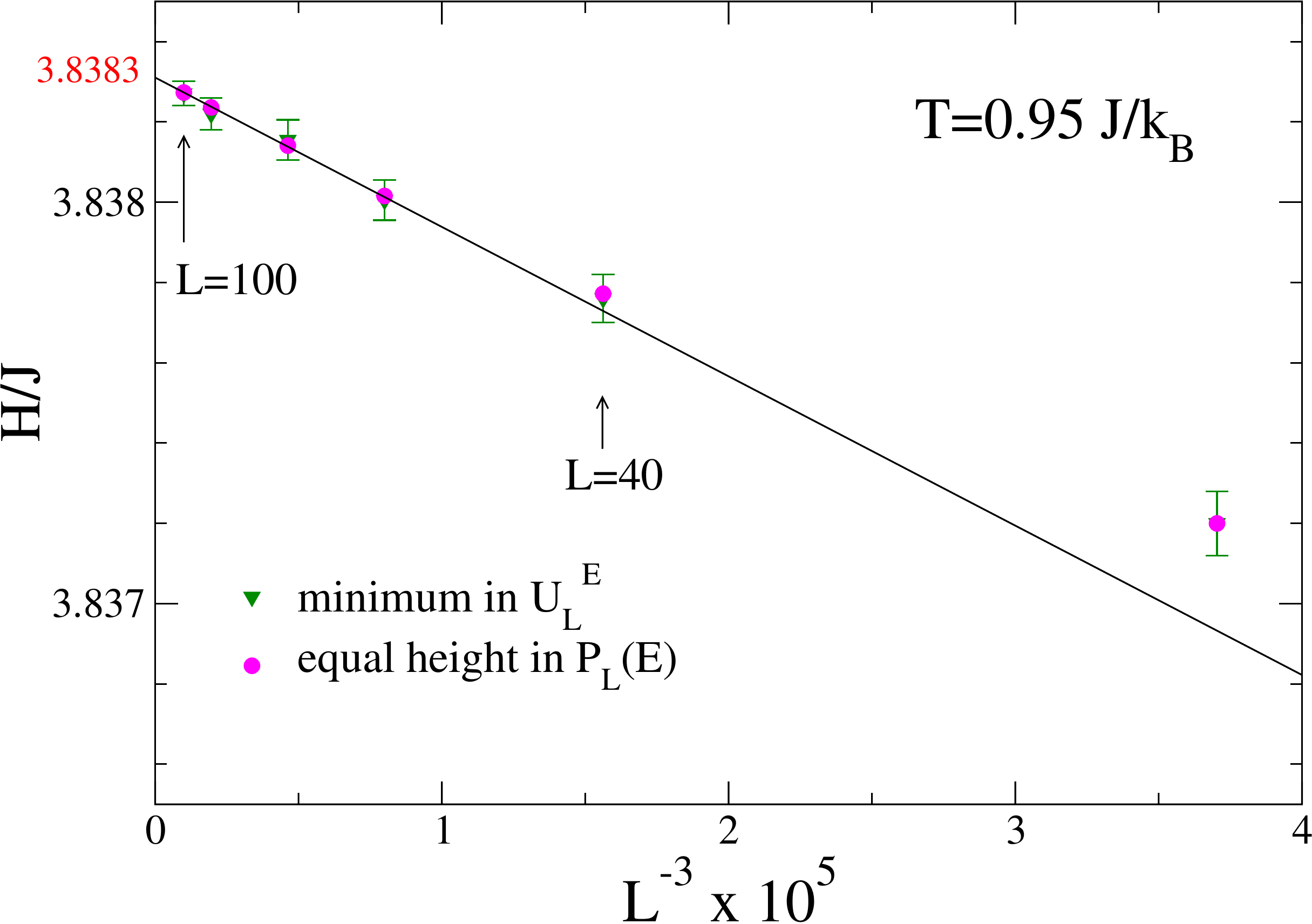}
\caption{\label{extrapEE} Extrapolation of the locations of the transition fields determined from the ``equal height'' rule for peaks in the probability distribution of the energy and from the minimum of the 4th order cumulant of the energy vs the inverse volume of the system.}
\end{figure}

The values of the transition field for each lattice size, $L$, as determined by the equal heights of the two peaks in the probability distributions for the energy, are plotted in Fig.~\ref{extrapEE}.  This figure shows very nicely that the variation is linear with $L^{-3}$ for $L \ge 40$.  The estimated transition field in the thermodynamic limit is $H^t/J=3.83830(5)$.   Also shown are the positions of the minima of the 4th order cumulant of the energy.  These agree almost perfectly with the values extracted from the locations of the peaks in the probability distributions and will be discussed in more detail later.

\begin{figure}
\centering
\includegraphics[clip,angle=270,width=0.95\hsize]{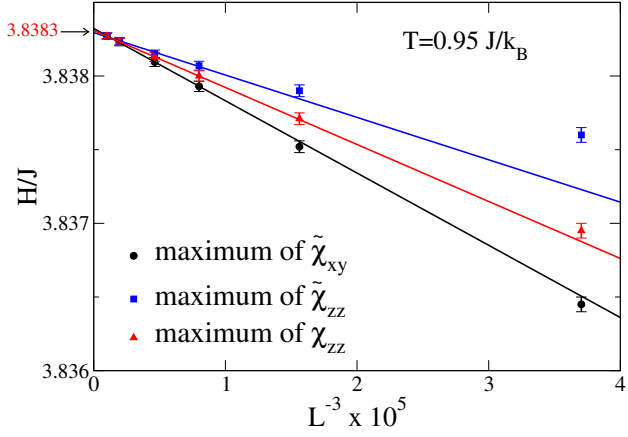}
\caption{\label{extrap} Size dependence of the finite size lattice transition field $H^t_L$ determined from the locations of the maxima of multiple susceptibilities vs $L^{-3}$ for lattices sizes from $L=30$ to $L=100$.  The solid lines show extrapolations to $L= \infty$ for $L \ge 50$.}
\end{figure}

From the extrapolation of the peak positions of $P_L(m_z)$ (and of the values of the first moments $\int dm_z m_zP_L(m_z)$ for each peak) we can estimate $m^{AF}_{z, \infty}$ and $m^{SF}_{z,\infty}$ and hence obtain the difference $\Delta m=m^{SF}_{z, \infty} - m^{AF}_{z, \infty} \approx 0.0352$. Since the factor $\Delta m L^3$ exceeds $k_BT/J$ by a factor from about $10^3$ to $1.8\times10^4$ when $L$ varies from $L=30$ to $L=80$, it is plausible that $H^t$ can be located with excellent precision, as shown in Figs.~\ref{extrapEE} and \ref{extrap}.  The fact that all extrapolations, using both quantities from the AF phase and from the SF phase, yield the same transition field $H^t$ to very good precision, reinforces the conclusion of Ref.~\cite{HTL} that there is a direct 1st order transition between the two phases with no intervening biconical phase.

\begin{figure}
\includegraphics[clip,angle=0,width=1.00\hsize]{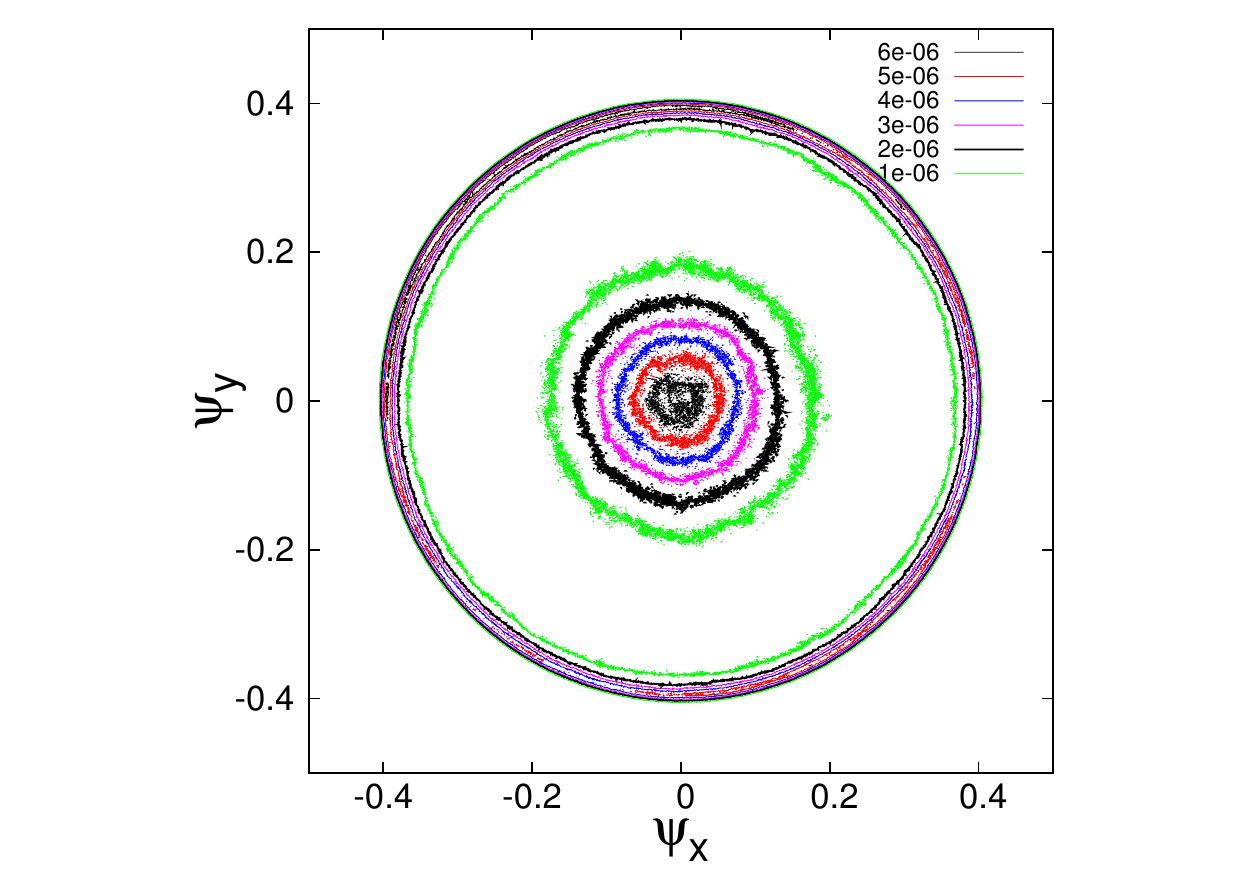}
\caption{\label{fig3} Contours for the order parameter distribution $P_L(\vec{\psi}) $ with $\vec{\psi}=(\psi_x, \psi_y$) being the two component order parameter comprising the $xy$-components of the staggered magnetization in the spin-flop phase for an $L=60$ lattice at $H = H^t_L$. Different colors (in the electronic version) denote the magnitude of the probability (from the center outwards the probability first decreases and then increases again).} 
\end{figure}

As a check on the assumptions about the degeneracy of the SF order parameter, in Fig.~\ref{fig3} we show the order parameter distribution in the xy-plane at the transition field $H^t_L$ for an $L=60$ lattice.  The contours of constant absolute value are almost perfectly circular and show a clear jump from the non-zero value in the SF-phase to a small value in the AF-phase which differs from zero only because of finite size effects.

The variation of the positions of the peaks in the susceptibilities of the uniform magnetization as well as both the z-component of the staggered magnetization and of the SF-order parameter are shown in Fig.~\ref{extrap}.  Excluding the values for $L=30, 40$ as probably being outside the asymptotic region, we fitted the remaining values to obtain an asymptotic value.  The positions of all three susceptibilities extrapolate with $L^{-3}$ to a value of $H^t/J=3.83830(8)$.  This is in perfect agreement with the result of the extrapolations presented in Fig.~\ref{extrapEE}.  Using the slope of the susceptibility for $m_z$ and Eq.~(\ref{eq13a}) we estimate an effective value $q\approx 3.7$ which is slightly larger than the estimate of $\pi$ as suggested in Sec.II.C.2.

\begin{figure}
\centering
\includegraphics[clip,angle=270,width=1.0\hsize]{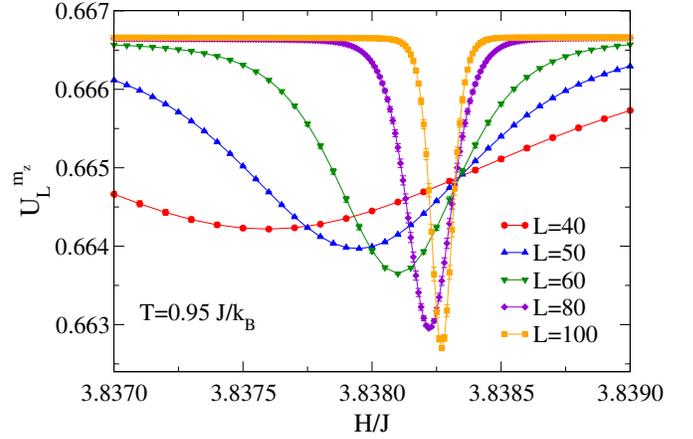}
\caption{\label{cumMz} Variation of the 4th order cumulant of the z-component of the magnetization vs $H$ for different lattice sizes.}
\end{figure}

The behavior of the 4th order cumulant of the z-component of the magnetization, $m_z$, seen in Fig.~\ref{cumMz}, shows a clear minimum for each lattice size which sharpens and moves towards larger fields as the lattice size increases.  With increasing $L$ the value of the minimum decreases.
For $L \ge 50$ the positions of these minima extrapolate linearly with $L^{-3}$ to a value of $H^t/J=3.83831(5)$ in excellent agreement with the extrapolated value obtained from the location of the transition from the ``equal weight'' rule.

\begin{figure}
\centering
\includegraphics[clip,angle=270,width=0.95\hsize]{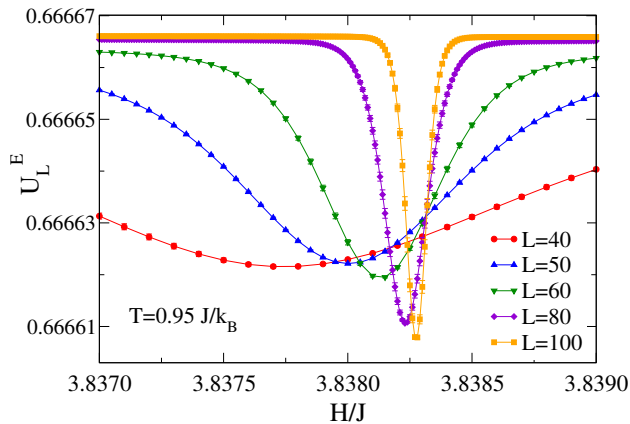}
\caption{\label{cumE} Variations of the 4th order cumulant of the energy for different lattice sizes vs applied field $H$.}
\end{figure}

The 4th order cumulant of the energy, seen in Fig.~\ref{cumE} also shows a single minimum that sharpens and moves slowly towards higher fields and becomes deeper as the lattice size increases.  For $L \ge 40$ the positions of these minima extrapolate linearly with $L^{-3}$ to a value of $H^t/J=3.83831(5)$, in excellent agreement with the behavior of the z-component of the magnetization (both the ``equal weight'' rule and the cumulants).

\begin{figure}
\centering
\includegraphics[clip,angle=270,width=0.95\hsize]{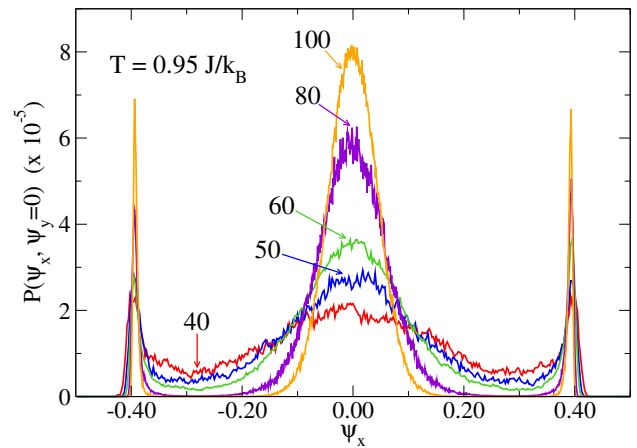}
\caption{\label{probMxys} Probability distribution (unnormalized) of the SF order parameter $\psi_x$, with $\psi_y=0$ for different lattice sizes.}
\end{figure}

The probability distributions of the SF order parameter $\psi$ at $T=0.95J/k_B$ are shown at the transition field for different lattice sizes $L$ in Fig.~\ref{probMxys}.  For each size we chose the finite size transition field to be located at the same point at which the (symmetric) peaks in the probability distribution for the energy were of equal heights.  These data show that the distributions contain two clear peaks at $\pm{\psi_{\infty}}$ corresponding to the SF order and a peak centered about zero corresponding to  the AF phase.  These peaks cannot be described solely by Gaussians since the states describing phase coexistence (from about $|\vec\psi|\approx 0.2$ to $|\vec\psi|\approx 0.36$) are not yet strongly suppressed.  If we ``separate'' the distributions into two peaks by choosing the minimum probability as the separation point, we can measure the ``weight'' of each peak by numerically integrating the probability under each peak.  As shown in Table~\ref{table_q_eff} the relative weight of the sum of the ``ordering'' peaks and the disordered peak  depends upon the exact choice of $H^t$ and is also slightly dependent upon the choice of $L$.  For our best estimate of $H^t/J=3.838305$ the value appears to be converging for large $L$ at the estimate of $q=\pi$ that was obtained earlier using the two Gaussian approximation.  Note that the result $q=\pi$ was already predicted from the two delta-function distribution appropriate to the thermodynamic limit in Eq.~(\ref{coex'}). 

\begin{table}[]
\caption {Estimates for $q_{eff}$ from the ratio of probability distributions of the weights of the peaks for different values of $L$.}
\label{table_q_eff}
\begin{ruledtabular}
\begin{tabular}{c c c c}
\colrule
$L$ & $H^t/J=3.83830$ & $H^t/J=3.838305$ & $H^t/J=3.83831$ \\
\colrule
60  & $3.31(23)$ & $3.42(30)$ & $3.53(28)$ \\
80  & $3.15(24)$ & $3.36(26)$ & $3.75(29)$ \\
100  & $2.82(30)$ & $3.19(29)$ & $3.82(30)$ \\
\end{tabular}
\end{ruledtabular}
\end{table}

\begin{figure}
\centering
\includegraphics[clip,angle=270,width=0.95\hsize]{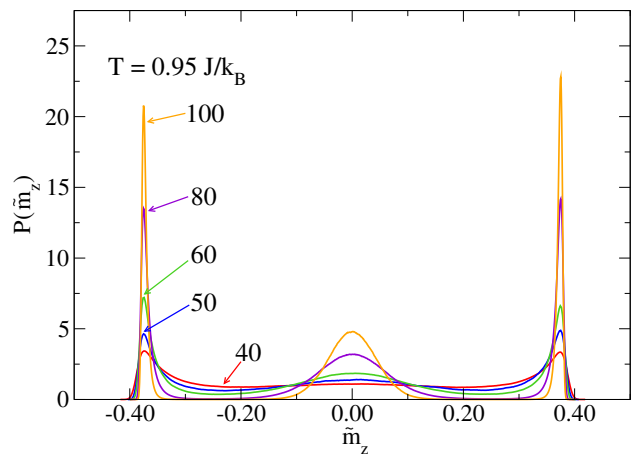}
\caption{\label{probMzs} Probability distribution (unnormalized) of the AF order parameter $\tilde{m_z}$, for different lattice sizes.}
\end{figure}

For completeness we show the probability of the AF order parameter vs the applied field in Fig.~\ref{probMzs}.  For small systems two peaks are seen at $\pm\tilde{m}_{\infty}$ with a broad plateau in between, but as $L$ increases three distinct peaks develop.  One peak, centered about $\tilde{m}_z=0$ is for the SF phase with no AF order and the peaks at $\pm\tilde{m}_{\infty}$ are for the AF ordered phase.  Again, the minima between the peaks for $L \ge 50$ can be interpreted in terms of phase coexistence.

\begin{figure}
\centering
\includegraphics[clip,angle=270,width=0.95\hsize]{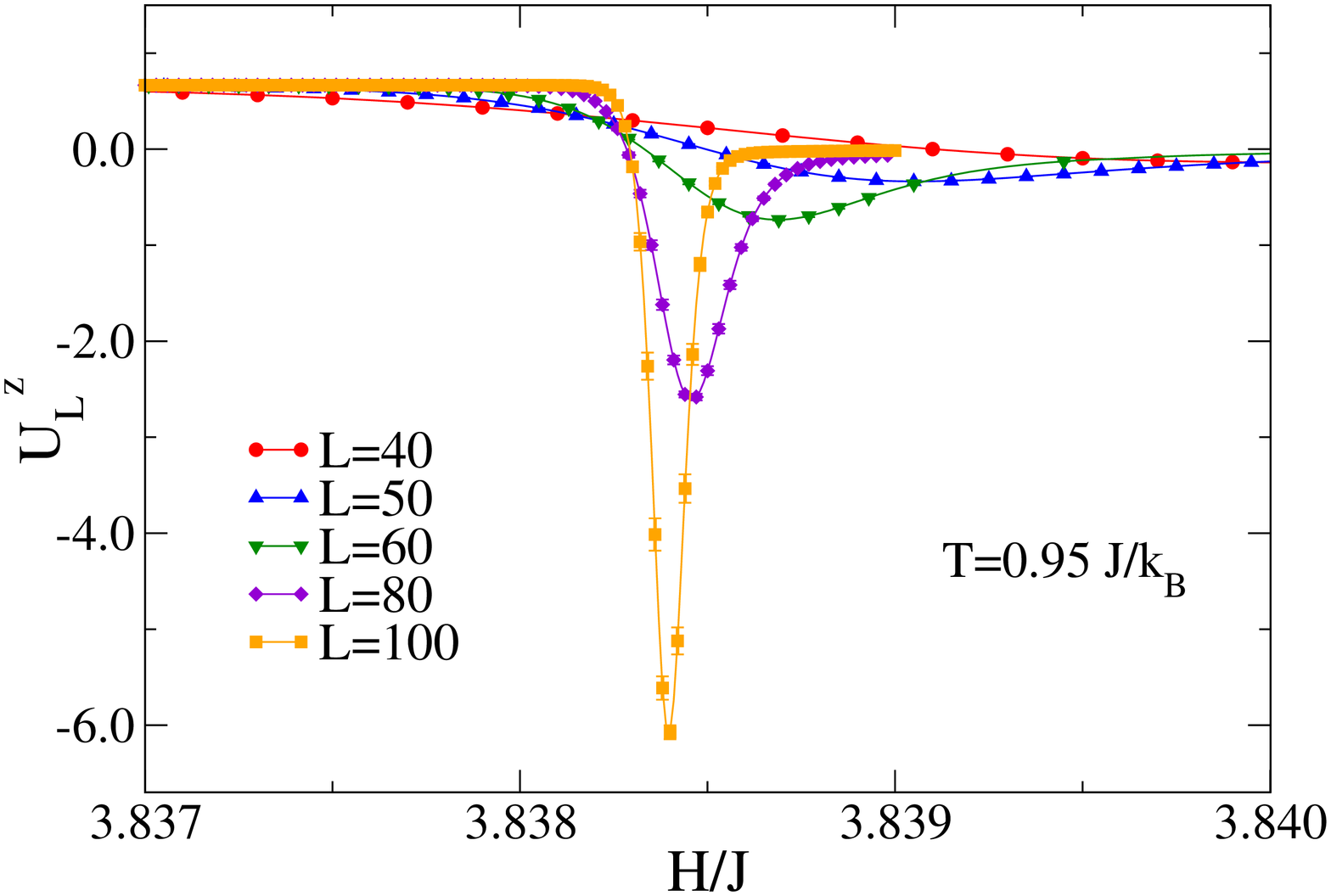}\vspace{0.3cm}
\includegraphics[clip,angle=270,width=0.95\hsize]{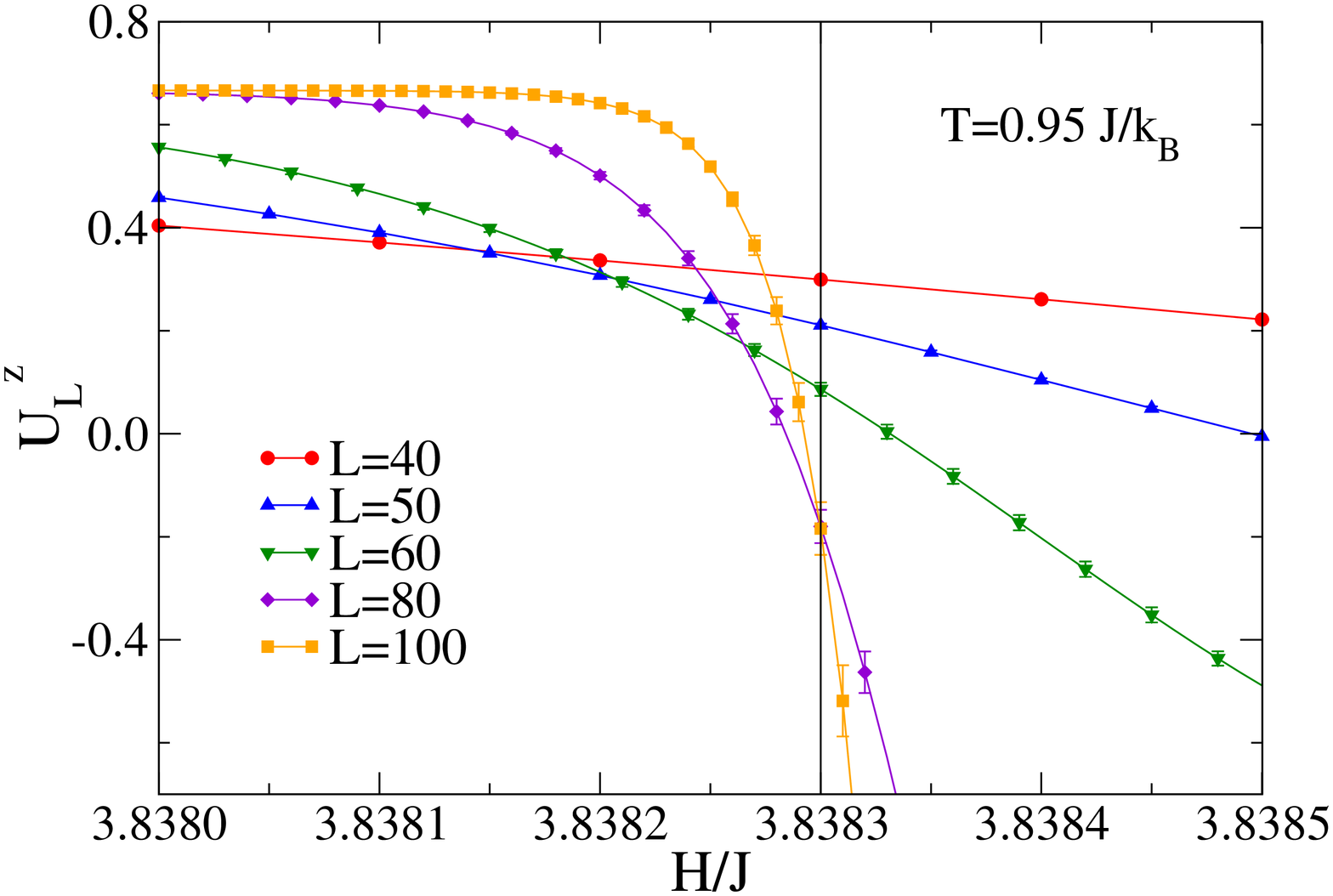}
\caption{\label{cumMzs} (top) Variation of the 4th order cumulant of the AF order parameter vs magnetic field for different lattice sizes; (bottom) Same as above but on a finer scale.}
\end{figure}

In Fig.~\ref{cumMzs} we show the variation of the 4th order cumulant of the z-component of the order parameter with field.  As the lattice size increases the crossing points move systematically towards slightly larger fields and the values of the cumulant at the crossing points decrease.  In fact, the reduction in the value of ${U}_L^z$ at the crossing seems to accelerate and there is no indication of convergence for the range of lattice sizes studied so from these data alone we cannot tell if the prediction from our simple phenomenological theory is verified.  However, in Appendix B we give a tentative interpretation of this behavior in terms of crossover behavior between critical behavior dominated by the bicritical point (prevailing for small $L$) towards first order finite size behavior.

\begin{figure}
\centering
\includegraphics[clip,angle=270,width=0.95\hsize]{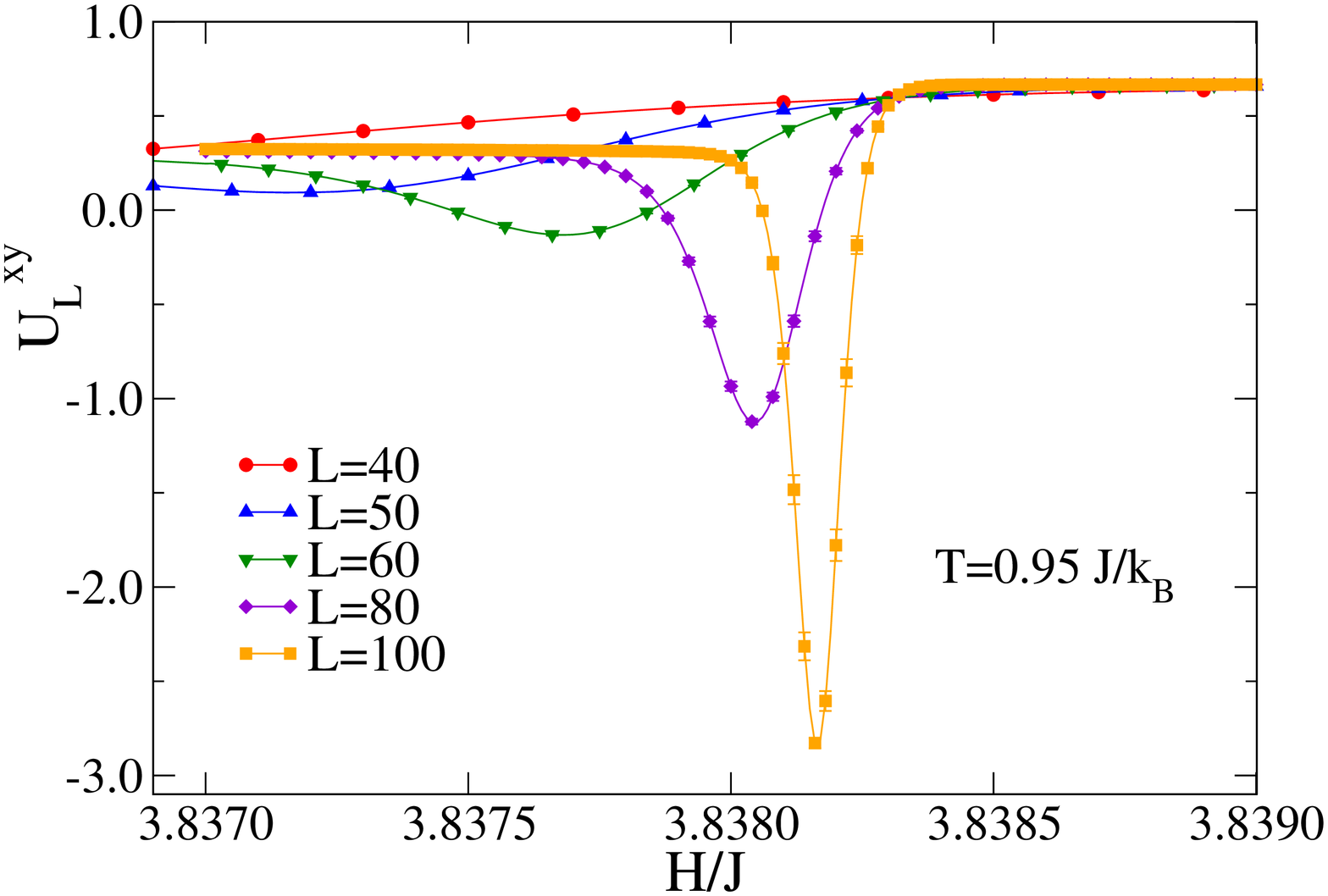}\vspace{0.3cm}
\includegraphics[clip,angle=270,width=0.95\hsize]{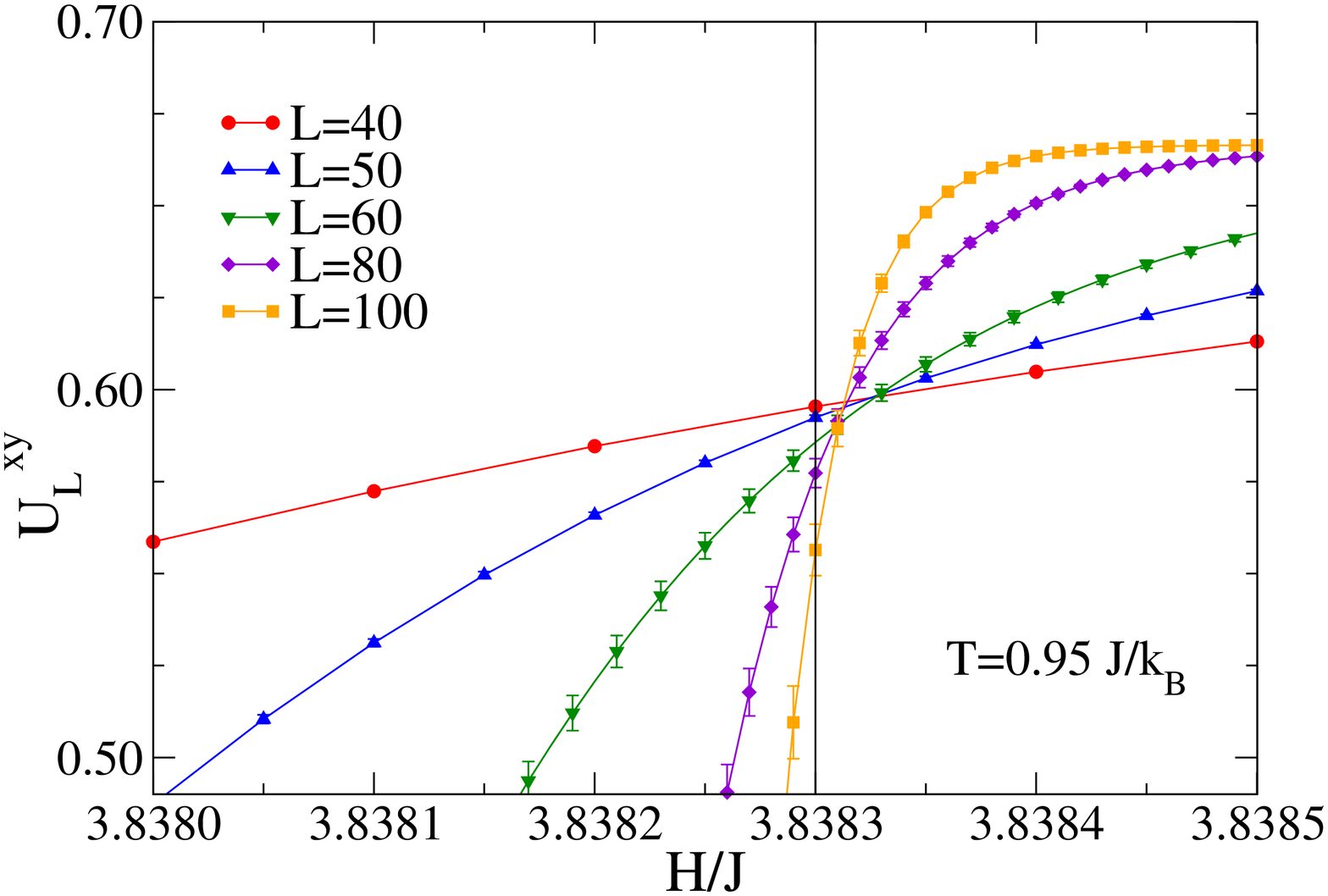}
\caption{\label{cumMxys} (top) Variation of the 4th order cumulant of the SF order parameter vs magnetic field for different lattice sizes; (bottom) Same as above but on a finer scale.}
\end{figure}

Data for the 4th order cumulant of the xy-component of the order parameter, shown in Fig.~\ref{cumMxys}, reveals similar behavior except that the minimum occurs at fields lower than the field of the crossing point and moves towards higher fields as the size increases.  The crossing point values are just below $U_{\infty}^{xy}=0.6$ and are higher than the predicted value of $U_{\infty}^{xy}=0.56056$ assuming an effective value of $q=\pi$.  However, the bottom portion of this figure shows a clear tendency for the crossing point values to decrease slightly with increasing size, and the predicted value is not inconsistent with an extrapolation to $L = \infty$.  (Note:  Use of $U_{\infty}^{xy}=0.6$ in Eq.~(\ref{eq32V}) would yield $q=5$).

\begin{figure}\vspace{0.5cm}
\centering
\includegraphics[clip,angle=270,width=0.95\hsize]{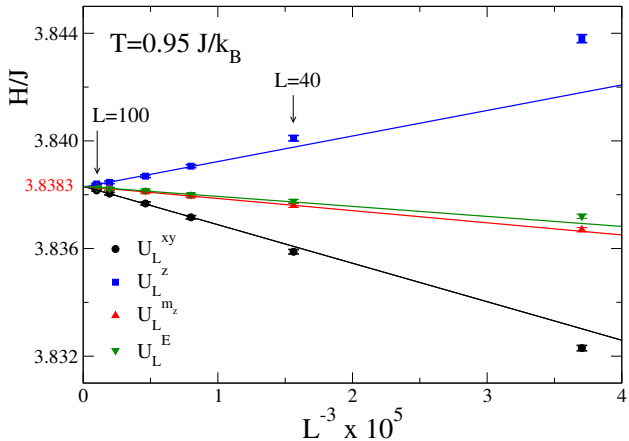}
\caption{\label{extrapmin} Variation of the minima in the 4th order cumulants of the AF and SF order parameters, the energy $E$ and the uniform magnetization $m_z$ for different lattice sizes.}
\end{figure}

  The locations of the minima for the cumulants of both order parameters as well as those of the energy and the uniform magnetization $m_z$ extrapolated to the thermodynamic limit are also shown in Fig.~\ref{extrapmin} vs the inverse volume, $L^{-3}$.  The observed variation with lattice size agrees with the predictions of our simple double Gaussian theory and, again, we find a common intersection point of $H^t/J=3.83830(7)$.  The asymptotic size regime, however, appears to begin only for $L\ge40$.  The observation that the depth of the minima decreases very strongly with increasing $L$ is compatible with a scaling as $\approx -L^3$ as observed in the Potts model~\cite{Vollmayr}.

\begin{figure}
\centering
\includegraphics[clip,angle=270,width=0.9\hsize]{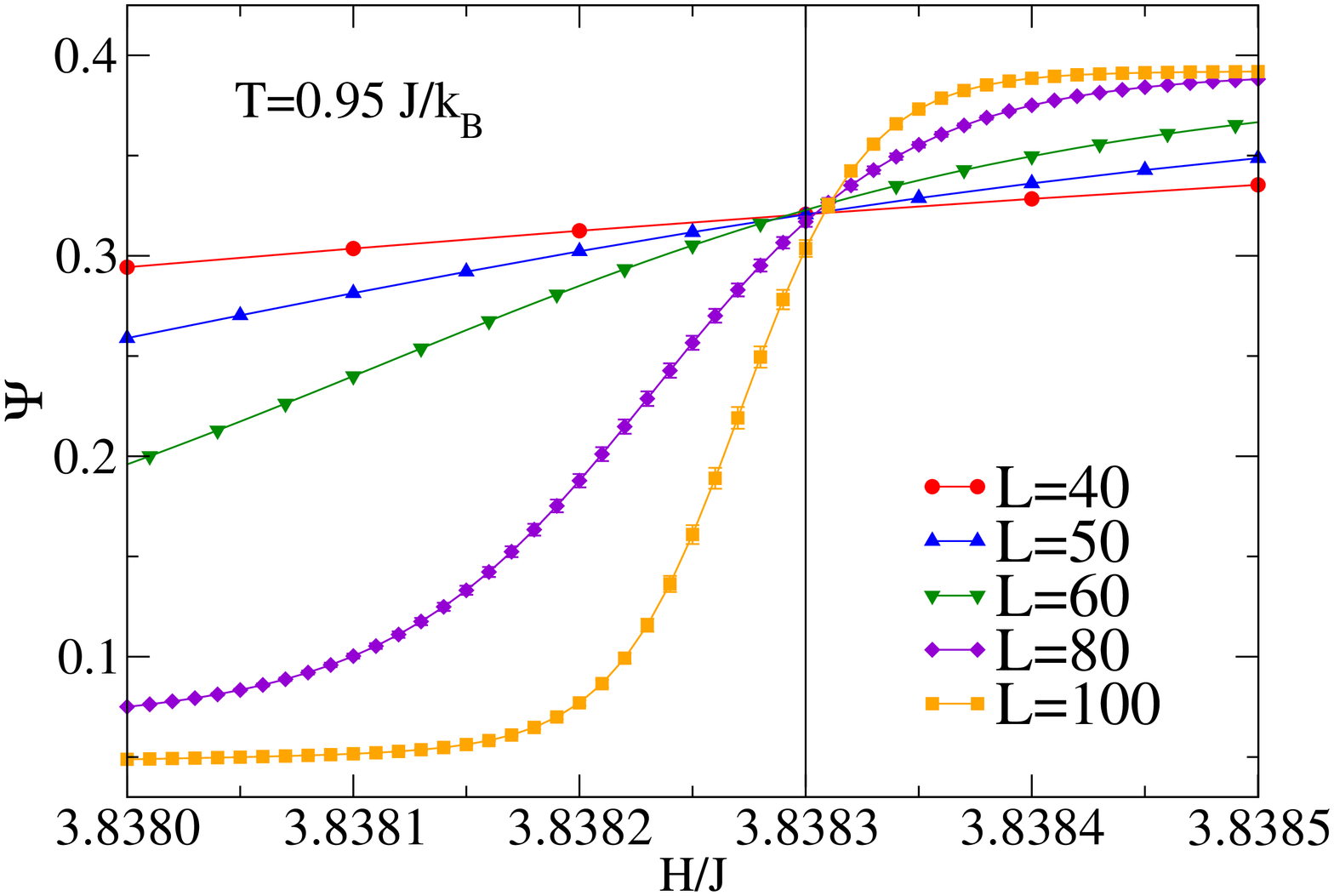}\vspace{0.5cm}
\includegraphics[clip,angle=270,width=0.9\hsize]{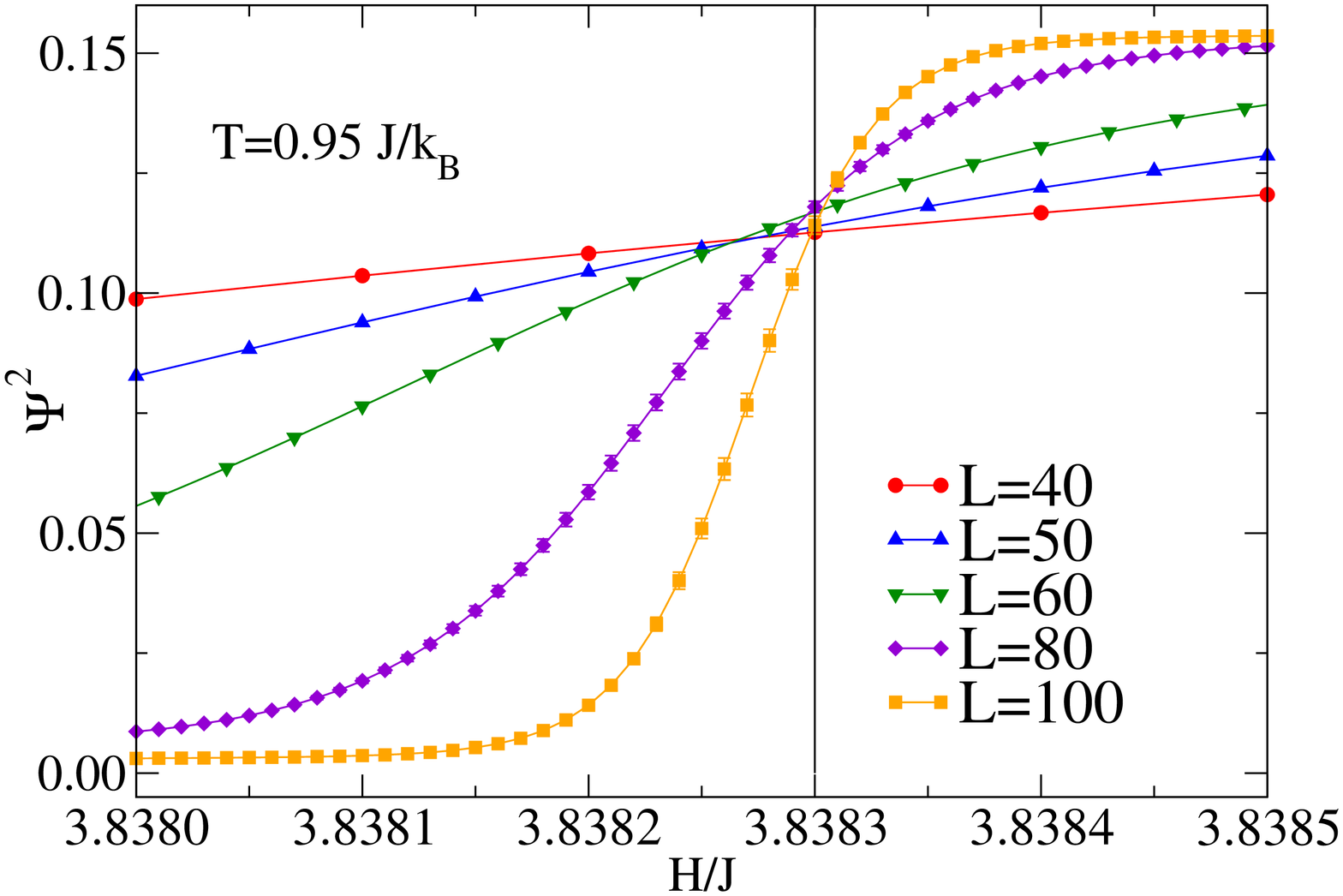}
\caption{\label{phiphi2} (top) Variation of the SF order parameter vs magnetic field for different lattice sizes; (bottom) Variation of the square of the xy-component of the order parameter.}
\end{figure}

\begin{figure}
\centering
\includegraphics[clip,angle=270,width=0.9\hsize]{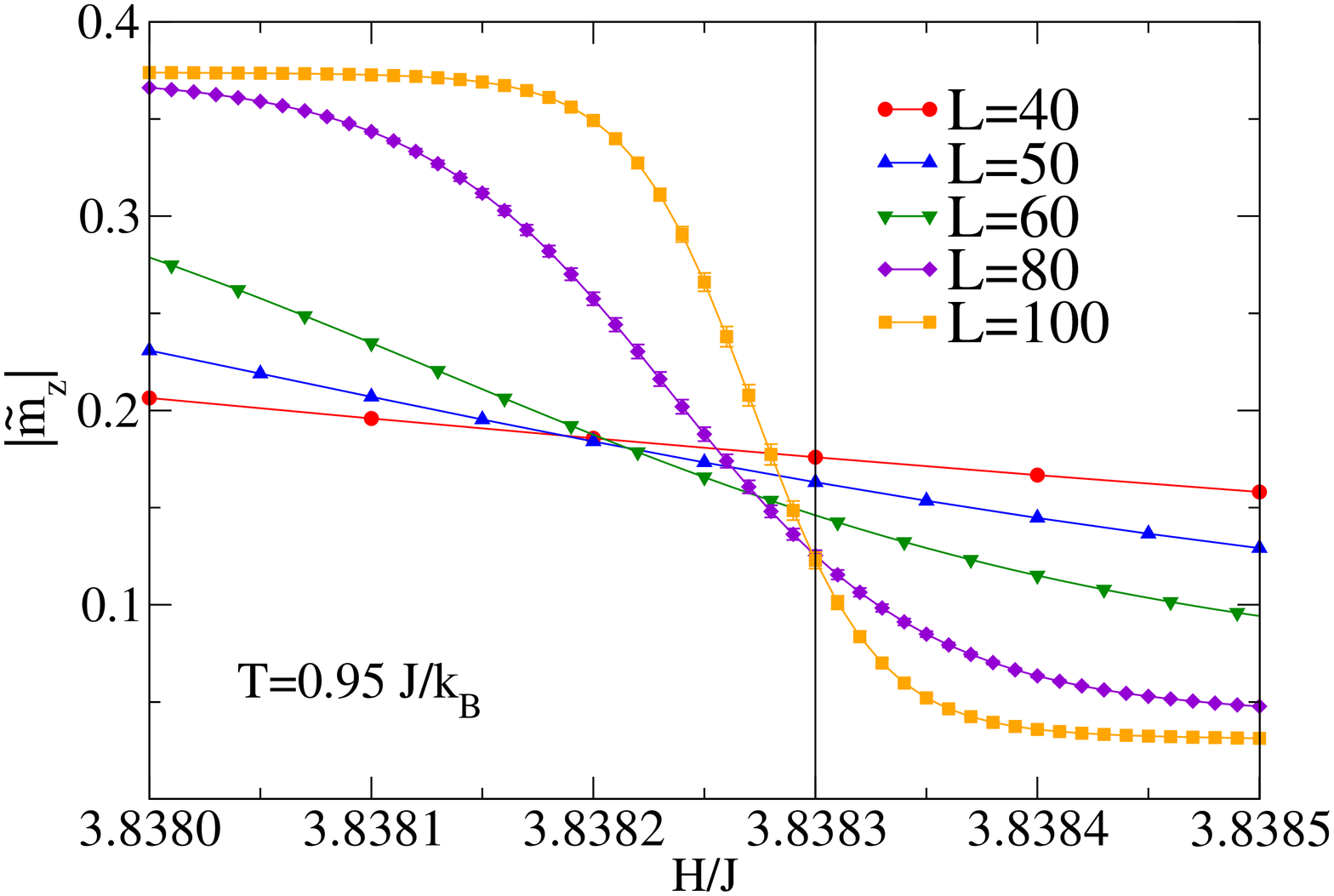}\vspace{0.5cm}
\includegraphics[clip,angle=270,width=0.9\hsize]{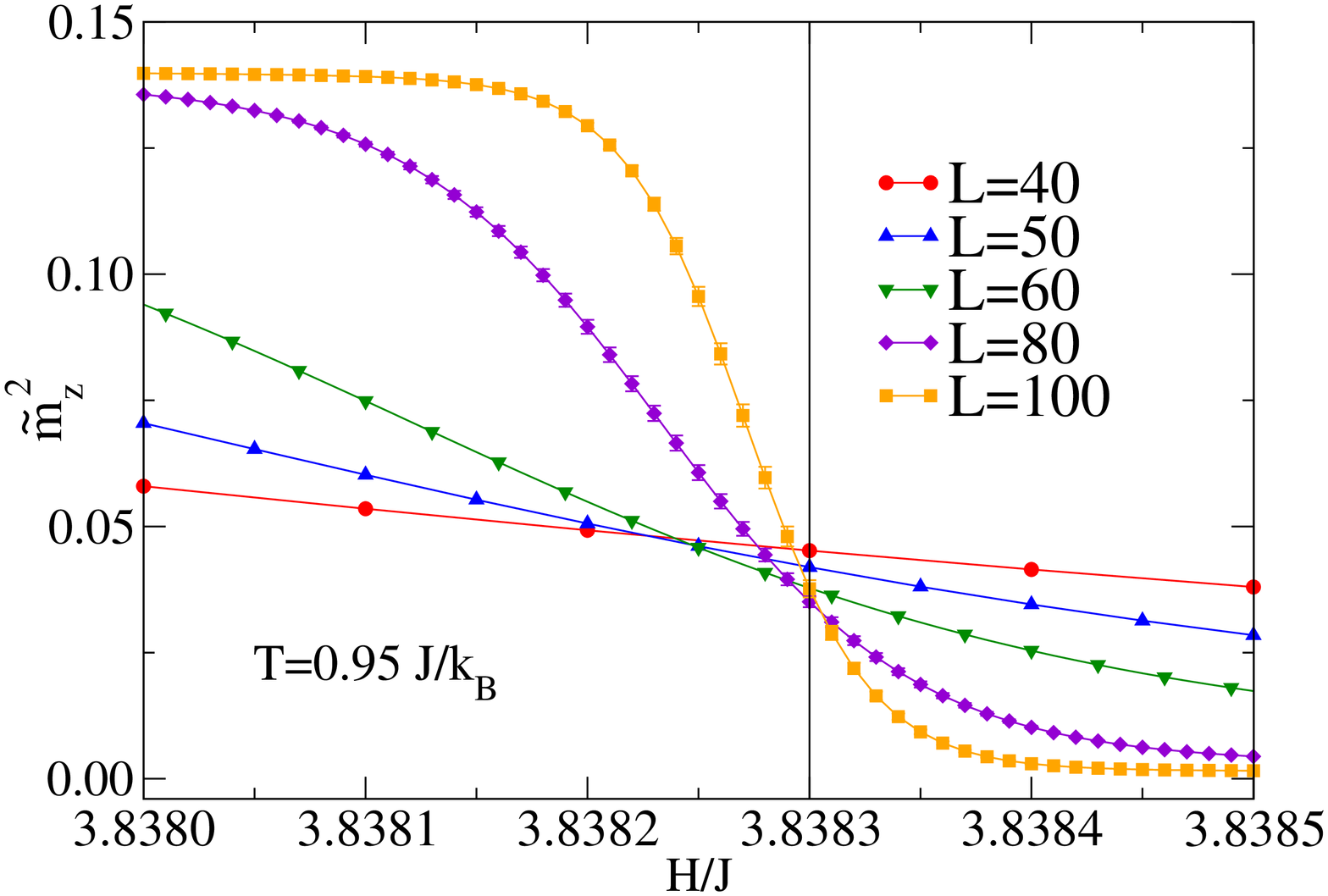}
\caption{\label{msms2} (top) Variation of the AF order parameter vs magnetic field for different lattice sizes; (bottom) Variation of the square of the z-component of the order parameter.}
\end{figure}

\begin{figure}
\centering
\includegraphics[clip,angle=270,width=0.9\hsize]{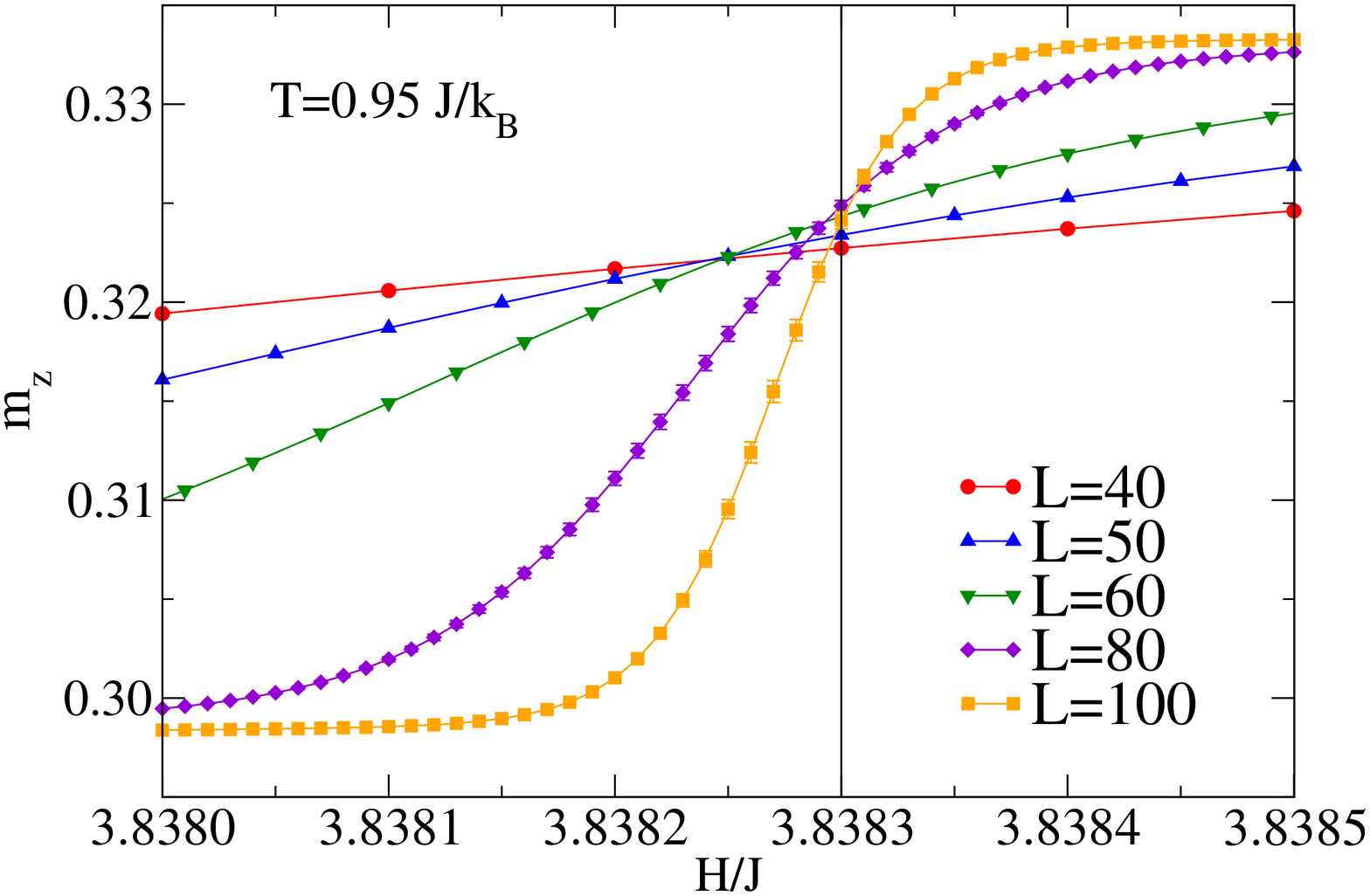}\vspace{0.5cm}
\includegraphics[clip,angle=270,width=0.9\hsize]{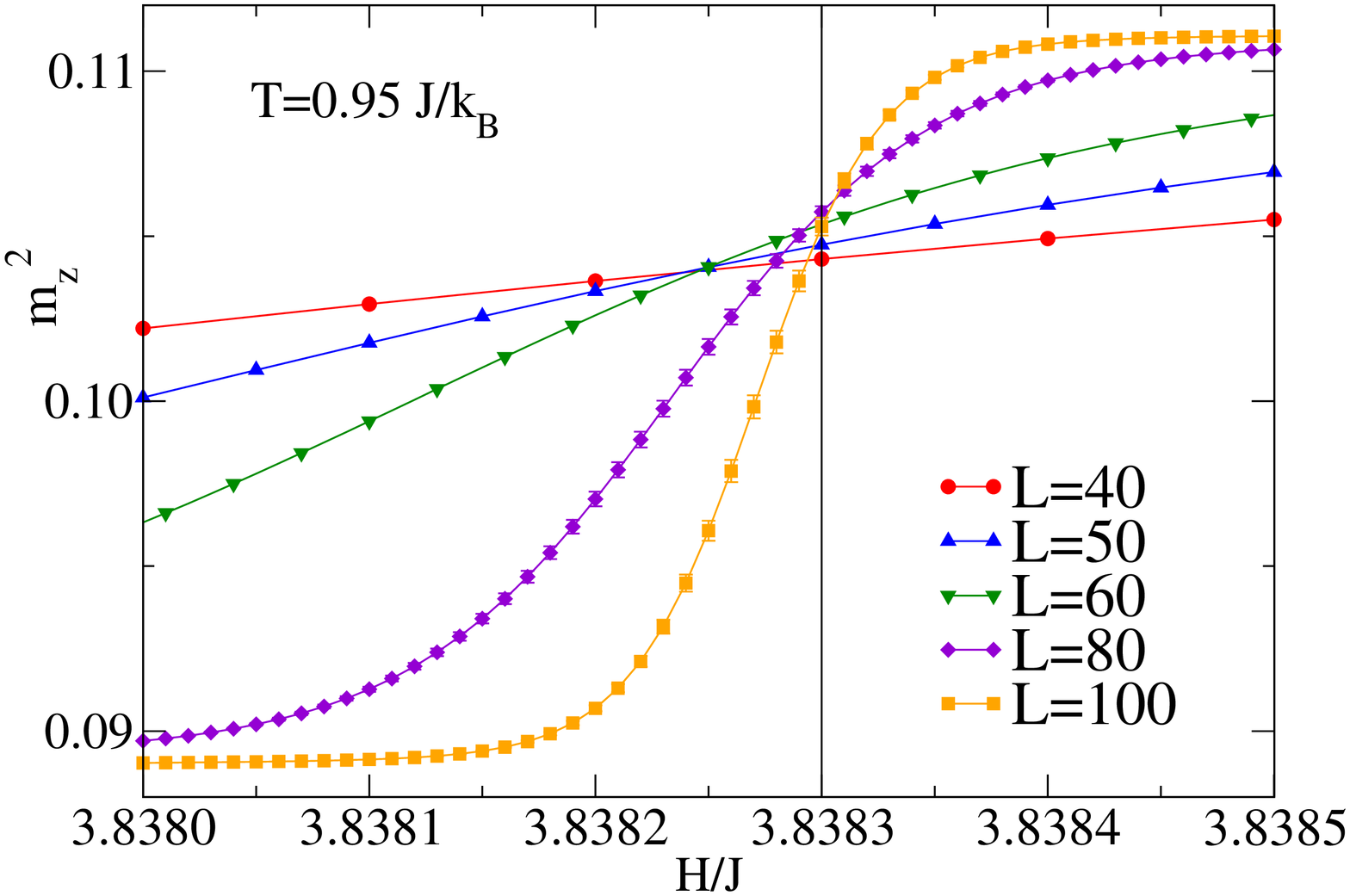}
\caption{\label{mm2} (top) Variation of the z-component of the magnetization vs magnetic field for different lattice sizes; (bottom) Variation of the square of the z-component of the magnetization.}
\end{figure}

 Perhaps the most striking results of our study emanate from Eqs. (16), (17), (19) and (20).  Plots of ${\langle|\vec{\psi}|\rangle}_L$, ${\langle\psi^2\rangle}_L$, ${\langle|\tilde{m}_z|\rangle}_L$, and ${\langle\tilde{m}_z^2\rangle}_L$, vs $H$ should show common intersection points for different $L$ at $H^t$.  These features would $\bf not$ occur at a second order transition, but are nicely consistent with our phenomenological theory.  In Figs. 16, 17, and 18 we show data for the first and second moments of $\psi$, $\tilde{m}_z$ and, for completeness, $m_z$ near the transition.
 From Eqs. (16) and (17) we conclude that
 
 \begin{equation}
\frac{ \langle|\vec{\psi}|\rangle_L|_{H^t}}{\psi_{\infty}}=\frac{\langle\psi^2\rangle_L|_{H^t}}{\psi_{\infty}^2}=\frac{q}{1+q}\approx 0.7585
 \end{equation}
 
 \noindent and from Eqs. (19) and (20)
 
 \begin{equation}
\frac{\langle|\tilde{m}_z|\rangle_L|_{H^t}}{\tilde{m}_{z,\infty}}=\frac{\langle\tilde{m}_{z}^2\rangle_L|_{H^t}}{\tilde{m}_{z,\infty}^2}=\frac{1}{1+q}\approx 0.2415 \enskip .
 \end{equation}
  
\noindent Here too, the data show small but systematic shifts with increasing system size and we can only say that the predictions are consistent with the current data which yield 
 
 \begin{eqnarray}
&&\frac{ \langle|\vec\psi|\rangle_L|_{H^t}}{\psi_{\infty}} \approx 0.77(3) \nonumber \\
&&\frac{\langle\psi^2\rangle_L|_{H^t}}{\psi_{\infty}^2} \approx 0.74(3) \nonumber \\
&&\frac{\langle|\tilde{m}_z|\rangle_L|_{H^t}}{\tilde{m}_{z,\infty}} \approx 0.32(4) \nonumber \\
&&\frac{\langle\tilde{m}_{z}^2\rangle_L|_{H^t}}{\tilde{m}_{z,\infty}^2} \approx 0.26  (3)\nonumber \enskip .
 \end{eqnarray}
  
 \noindent These values are in quite reasonable agreement with predictions although more precise values would be needed to draw strong conclusions.  However, the discrepancies between the measured and predicted values noted above can probably be attributed to the difference in the location of the intersections and our best estimate for $H^t/J=3.83830$.  Using more precise data on still larger systems to extrapolate the small finite size variations to $L\to \infty$ could give slightly different estimates than quoted above but would require prohibitively large resources at the present time.  
 
\begin{figure}
\centering
\includegraphics[clip,angle=270,width=0.9\hsize]{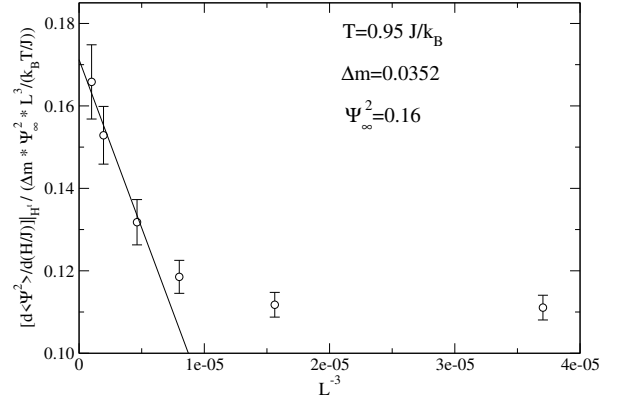}
\caption{\label{maxderiv} Variation of the `normalized' slope of $\langle\psi^2\rangle_L$ with respect to $H$ at the transition field $H^t$ vs the inverse volume $L^{-3}$ of the system.  The straight line is a linear extrapolation to the thermodynamic limit.}
\end{figure}

An alternative approach is to base our analysis on Eq.~(\ref{eq3333}) which describes the slope of $\langle\psi^2\rangle_L$ at the transition field $H^t$ as a function of $H/k_BT$.  From the corresponding plot of $d\langle\psi^2\rangle_L/d(H/J)|_{H^t}$, normalized by $J\Delta{m}L^3\psi_{\infty}^2/k_BT$, vs $L^{-3}$ we expect linear behavior whose intercept is $q/(1+q)^2 \approx 0.182$ if $q=\pi$.  Such a plot, shown in Fig.~\ref{maxderiv} is, indeed, reasonably compatible with this conjecture.
   
\begin{figure}
\centering
\includegraphics[clip,angle=270,width=0.95\hsize]{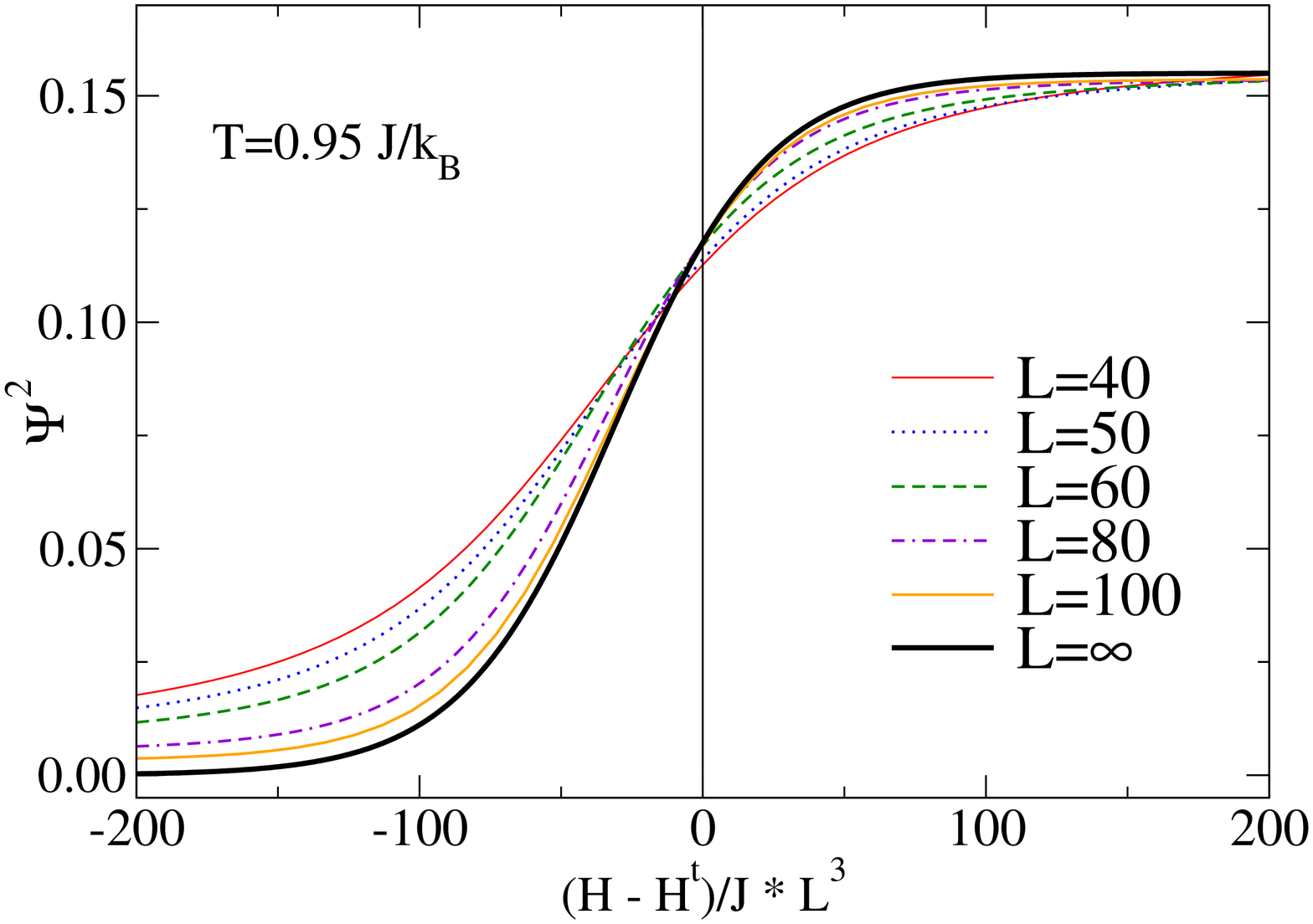}
\caption{\label{fsspsi2} Finite size scaling plot for $\langle\psi^2\rangle_L$.   The heavy, solid curve shows the asymptotic theoretical behavior in the limit $L = \infty$.}
\end{figure}

Near the transition, where $b$ can be neglected, Eq.~(\ref{eq100}) leads to a simple finite size scaling expression for  $\psi^2$ that is tested in Fig.~\ref{fsspsi2} in which $\psi^2$ is plotted vs. $(H-H^t)L^3$ where $H^t/J=3.83830$.  For large enough values of $L$ and $H-H^t$, the values for $\psi^2$ collapse onto a single curve representing a simple analytic scaling function describing the behavior near the transition between these two different ordered phases.  For large but negative values of $H-H^t$, curves should instead approach the small constant $b$ in Eq.~(\ref{eq100}) which goes to zero as $L^{-3}$.   The data bear out these predictions.

\section{Conclusion}
In order to provide an understanding of the finite size behavior of a first order transition from a state with simple, discrete degeneracy to a state with an infinite degeneracy, we have performed both theoretical and simulational studies of a uniaxially anisotropic Heisenberg antiferromagnet on finite, simple cubic lattices in an external field $H$ applied along the easy axis. 

We first presented a phenomenological theory based upon phase coexistence in the thermodynamic limit with probability distributions of the system in each phase described by delta functions.  We hypothesized that the relative weights of the AF and SF phases are $2$ and $2\pi$ by integrating over the angle $\phi$ of the two-component SF order parameter.  For finite volume, this description was generalized in terms of suitable Gaussian distributions, which led to the prediction that the moments and cumulants of the order parameters of both phases show common intersection points for large $L$ at the transition field $H^t$ (apart from corrections of order $L^{-3}$).  The values predicted for these intersections depend upon the effective, relative degeneracy $q=\pi$.  We then tested these somewhat speculative predictions via large scale Monte Carlo simulations

We determined the finite size behavior of the model by performing high resolution Monte Carlo simulations at $T = 0.95J/k_B$.  The phase transition can be located quite precisely by using an equal height rule for the probability distribution for the internal energy, and we find that an "equal weight" rule applies for the order parameter at the transition.  The asymptotic finite size behavior of the transition field as well as other quantities does not appear until lattice sizes $L\ge40$ are reached.  In retrospect this is not surprising since the discontinuity in the z-component of the magnetization is quite small, i.e. $\Delta{m_z}\le0.04$.  The locations of the minima in the cumulants for the antiferromagnetic order, spin-flop order, and internal energy extrapolate to the same transition field in the thermodynamic limit as does the ``equal weight'' rule for the magnetization as predicted by the double Gaussian approximation.  Different predictions from the theory yield consistent values for the effective value of the degeneracy $q$ but $\bf only$ for quite large values of $L$.    Therefore,  we conclude that the simple theory based upon a double Gaussian distribution provides a complete picture of the finite size effects at first order transitions between phases with different symmetries of the order parameters.  Since the underlying theory does not depend on the fine characteristics of the model, this means that a heretofore unknown kind of universality at a first order transition has been identified. \\

\begin{acknowledgments}
DPL thanks the Graduiertenschule MAterials science IN mainZ (MAINZ) and the Alexander von Humboldt Foundation for support. KB thanks the University of Georgia 's Office of the Vice President for Research at the U. of Georgia for support.  This work was supported in part 
by resources from the Georgia Advanced Computing 
Resource Center, a partnership between the University of Georgia's Office 
of the Vice President for Research and Office of the Vice President for 
Information Technology.
\end{acknowledgments}

\appendix 
\section{}
In this Appendix we want to provide some of the details of the calculations behind the phenomenological theory presented in Sec. II.  

The distribution of the order parameter $\vec\psi$ in the antiferromagnetic phase, given by Eq.~(\ref{eq14}) in the main text,

\begin{equation} \label{eqA1}
P^{AF}_L (\vec{\psi}) = \mathcal {N} \exp \Big(-\frac{{\vec\psi} {{\:}^2}}{2 k_BT \tilde{\chi}^{AF}_{xy} /L^3} \Big) \enskip,
\end{equation} 

\noindent can be used to evaluate the 4th order cumulant in the following way.
The normalization factor $\mathcal{N}$ is given by,

\begin{eqnarray} \label{eq15}
&& \mathcal{N}^{-1}=2 \pi \int\limits^\infty_0 \psi d \psi \exp(-\psi^2 L^3/(2k_BT \tilde{\chi}^{AF}_{xy})) \nonumber\\
&&\qquad=2 k_BT \pi \tilde{\chi}^{AF}_{xy} /L^3 \enskip .
\end{eqnarray}

\noindent This factor $\mathcal{N}$ is only applicable for the order parameter moments of a ``pure'' AF phase and has nothing to do with the normalization factor used to derive Eq.~(\ref{eq9}).  The second moment is simply

\begin{equation} \label{eq16}
\langle \psi^2 \rangle_{AF} = 2 k_BT \tilde{\chi}^{AF}_{xy} /L^3
\end{equation}

\noindent and the fourth moment

\begin{equation} \label{eq17}
\langle \psi ^4 \rangle _{AF} =2 ( 2 k_BT \tilde{\chi}^{AF}_{xy} /L^3 )^2 \enskip.
\end{equation}

\noindent Hence

\begin{equation} \label{eq18}
U^{AF}_L= 1- \langle \psi^4 \rangle_{AF}/[3 \langle \psi^2 \rangle^2_{AF} ]=1/3
\end{equation}

\noindent Note that for a two-component order parameter, a different normalization of $U_L$ would be required for a Gaussian distribution to yield zero in the disordered phase.

In order to obtain the moments of the order parameter of the SF phase using Eq.~(\ref{eq19}), we have to compute the integrals

\begin{equation} \label{eq20}
\langle \psi^2 \rangle_{SF} = 2 \pi\mathcal{N} \int\limits^\infty_0 \psi d \psi \, \psi^2 \exp \Big[-\frac{(\psi^2 - \psi^2_\infty)^2 L^3}{8\psi^2_\infty k_BT \tilde{\chi}^{SF}_{xy}} \Big] \enskip ,
\end{equation}

\begin{equation} \label{eq21}
\langle \psi ^4 \rangle_{SF} = 2\pi\mathcal{N} \int\limits^\infty_0 \psi d \psi  \psi^4 \exp \Big[- \frac{(\psi^2-\psi^2_\infty)^2 L^3}{8\psi^2_\infty k_BT \tilde{\chi}
^{SF}_{xy}}\Big] \enskip,
\end{equation}

\noindent where the normalization factor $\mathcal{N}$ is given by

\begin{eqnarray} \label{eq22}
&&\mathcal{N}^{-1} = 2 \pi \int\limits^\infty_0 \psi d \psi \exp \Big[- \frac{(\psi^2 - \psi^2_\infty)^2 L^3}{8\psi^2_\infty k_BT \tilde{\chi}^{SF}_{xy}} \Big] \nonumber\\
&&\qquad=2\psi_{\infty}\pi \sqrt{2 \pi k_BT \tilde{\chi}^{SF}_{xy}} L^{-3/2} \enskip.
\end{eqnarray}

 Similarly, writing $\psi^2=x$, we conclude

 \begin{equation} \label{eq26}
 \langle \psi^2 \rangle_{SF} =\pi\mathcal{N} \int\limits^\infty_0 dx x \, \exp\Big[-\frac{(x-\psi^2_\infty)^2L^3}{8\psi^2_\infty k_BT \tilde{\chi}^{SF}_{xy}}\Big] \approx \psi^2_\infty
 \end{equation}
 \\

\noindent  with negligibly small correction. However, a non-trivial correction term arises in the fourth moment,

\begin{eqnarray} \label{eq29}
&& \langle \psi^4 \rangle_{SF} =\pi\mathcal{N} \int\limits_0^\infty dx x^2 \, \exp \Big[-\frac{x^2 - 2x \psi^2_\infty + \psi^4_\infty}{8\psi^2_\infty k_BT \tilde{\chi}^{SF}_{xy} /L^3} \Big] \nonumber\\
&&\qquad\quad\approx \psi^4_\infty + 4\psi^2_\infty k_BT \tilde{\chi}^{SF}_{xy} /L^3 \enskip.
 \end{eqnarray}

The ratio $U_L$ in the ordered SF phase hence becomes

\begin{equation} \label{eq32}
U^{SF}_L= 1- \langle \psi ^4 \rangle_{SF}/(3 \langle \psi^2 \rangle^2_{SF}) = \frac{2}{3} - \frac{4k_BT \tilde{\chi}^{SF}_{xy}} {3 L^3 \psi ^2_\infty} \enskip .
\end{equation}

Invoking the superposition approximation of Eq.~(\ref{eq32a})-~(\ref{eq32a'}) we find that the 4th order cumulant for the SF order then becomes

\begin{widetext}
\begin{equation} \label{eq33}
U_L^{xy}=1- 
\frac{[1+q\exp(\mathcal{Z})][q\exp(\mathcal{Z})[1+ 4k_BT \tilde\chi_{xy}^{SF}/(\psi_\infty^2L^3)]+2(2k_BT\tilde\chi_{xy}^{AF}/(\psi_\infty^2L^3))^2]}{3[q\exp(\mathcal{Z})+2k_BT \tilde\chi_{xy}^{AF}/(\psi_\infty^2L^3)]^2} \enskip .
\end{equation}\\
\end{widetext}

Using Eqs.~(\ref{eq35'}) and ~(\ref{eq35''}) we can straightforwardly obtain the second and fourth moments $\langle\tilde{m}_z^2\rangle, \langle\tilde{m}_z^4\rangle$ in both phases and apply a superposition approximation.  The resulting cumulant is

\begin{widetext}
\begin{equation} \label{eq33a}
{U_L^z}=1- 
\frac{[1+q\exp(\mathcal{Z})]}{3}\frac{{[1+6k_BT\tilde\chi_{zz}^{AF}/(\tilde{m}_{z,\infty}^2L^3)+3(k_BT\tilde\chi_{zz}^{AF})^2/(\tilde{m}_{z,\infty}^2L^3)^2+3q\exp(\mathcal{Z})(k_BT\tilde\chi_{zz}^{SF})^2/(\tilde{m}_{z,\infty}^2L^3)^2)}]}{[1+k_BT \tilde\chi_{zz}^{AF}/(\tilde{m}_{z,\infty}^2L^3)+q\exp(\mathcal{Z})k_BT\tilde\chi_{zz}^{SF}/(\tilde{m}_{z,\infty}^2L^3)]^2} \enskip .
\end{equation}\\
\end{widetext}

The weighted averages of the order parameter moments in the AF and SF phases become

\begin{equation} \label{eq119}
\langle |\tilde{m}_z|\rangle_L = \frac{\tilde{m}_{z, \infty}} {[1 + q \exp (\mathcal{Z})]}
+ \frac{q \exp (\mathcal{Z})  \sqrt{2k_BT\tilde
\chi^{SF}_{zz}/({\pi}L^3)}}{[1 + q \exp (\mathcal{Z})]} \enskip,
\end{equation}

\noindent and

\begin{widetext}
\begin{equation} \label{eq1110}
\langle|\vec\psi|\rangle_L=\Big[\psi_\infty (1-k_BT\tilde\chi_{xy}^{SF}/(2L^3\psi_{\infty}^2))q \exp (\mathcal{Z})
+ {\sqrt{\pi k_BT \tilde{\chi}^{AF}_{xy}/({2}L^3)}}\Big]/[1 + q \exp (\mathcal{Z})] \enskip .
\end{equation}\\
\end{widetext}

\subsubsection{Staggered susceptibility maxima}
In the absence of symmetry breaking staggered fields, the staggered susceptibility components $\tilde{\chi}'_{zz}$ and $\tilde{\chi}'_{xy}$, referring to the z-component $\tilde{m}_z$ and the xy-components $\vec{\psi}$ of the staggered magnetization, are defined as follows

\begin{equation} \label{eq111a}
k_BT \tilde{\chi}'_{zz} = L^3 (\langle \tilde{m}^2_z \rangle_L - \langle |\tilde{m}_z|\rangle ^2_L) \enskip,
\end{equation}

\begin{equation} \label{eq112}
k_BT \tilde{\chi}'_{xy} =L^3 (\langle \psi^2 \rangle _L-\langle|\vec\psi|\rangle^2_L) \enskip.
\end{equation}

\noindent Note that Eq.(\ref{eq111a}) yields the usual staggered susceptibility in the AF phase, and Eq.~(\ref{eq112}) in the SF phase.  Of course, in the phases where no spontaneous order exists, we simply have from the standard fluctuation relations

\begin{equation} \label{eq113}
k_BT \tilde{\chi}_{zz} =L^3 \langle \tilde{m}^2_z \rangle_L \enskip , \quad  SF \, {\rm phase} \enskip ,
\end{equation}

\begin{equation} \label{eq114}
k_BT \tilde{\chi}_{xy}=L^3 \langle \psi^2 \rangle_L \enskip , \quad AF\, {\rm  phase} \enskip.
\end{equation}

It is well known that a unique expression for the staggered susceptibility in both phases would require consideration of the limit of an applied staggered field (conjugate to the appropriate order parameter) approaching zero after the thermodynamic limit had been taken.  This, however, would be quite inconvenient within the context of simulations.

As already discussed above, the moments $\langle \cdots \rangle_L$ are computed from the corresponding moments $\langle \cdots \rangle_{AF},$ $ \langle \cdots \rangle _{SF}$ in the pure phases, taking an average with the corresponding weights $a_{AF}$, $1-a_{AF}$. For the second moments, this has already been considered above; here we supply the results for the first moments. For the AF phase we find

\begin{equation} \label{eq115}
\langle |\tilde{m}_z | \rangle _{AF} \approx \tilde{m}_{z,\infty} \enskip, \quad \langle \tilde{m}^2_z \rangle_{AF}\approx \tilde{m}^2_{z, \infty} +k_BT \tilde{\chi}^
{AF}_{zz}/L^3 \enskip ,
\end{equation}

\noindent which together with Eq.~(\ref{eq111a}) yields the correct result $\tilde{\chi}'_{zz} =\tilde{\chi}^{AF}_{zz}$ in the AF phase.  For the SF phase we find

\begin{equation} \label{eq116}
\langle |\vec\psi|\rangle_{SF} \approx \psi_\infty (1-k_BT\tilde\chi_{xy}^{SF}/(2L^3\psi_{\infty}^2)) \enskip , \quad \langle \psi^2 \rangle_{SF} \approx \psi^2_\infty \enskip.
\end{equation}\\

\noindent From Eqs.~(\ref{eq32a}),~(\ref{eq35'}),~(\ref{eq35''}),~(\ref{eq16}),~(\ref{eq26})~(\ref{eq119}), and~(\ref{eq1110}) it is straightforward to obtain the expressions for the staggered susceptibilities
$k_BT \tilde{\chi}'_{zz}$, $k_BT \tilde{\chi}'_{xy}$ defined in Eqs.~(\ref{eq111a}),~(\ref{eq112}). These expressions show nicely that in the AF phase, i.e. for $\mathcal{Z} \rightarrow -\infty$

 \begin{equation} \label{eq1111a}
 \tilde{\chi}^{'AF}_{zz} =\tilde{\chi}^{AF}_{zz} \enskip, \quad \tilde{\chi}^{'AF}_{xy} = 2\tilde{\chi}_{xy}^{AF} \Big( 1- \frac{\pi}{4}\Big) \enskip.
 \end{equation}

\noindent Recall that the AF phase plays the role of the disordered phase for the SF order. Likewise,

\begin{equation} \label{eq1112}
\tilde{\chi}^{'SF}_{zz} = \tilde{\chi}^{SF}_{zz} \Big( 1 -\frac{2}{\pi} \Big) \enskip, \quad \tilde{\chi}^{'SF}_{xy} = \tilde{\chi}^{SF}_{xy} \enskip,
\end{equation}

\noindent since here the SF phase plays the role of the disordered phase for AF order.

Locating the staggered susceptibility maximum and estimating its height is also an interesting task. Using the abbreviation $q \exp (\mathcal{Z}) = Y$, we find from $d(\langle
\tilde{m}^2_z \rangle - \langle |\tilde{m}_z|\rangle^2) /dY = 0$ that the maximum occurs for

\begin{widetext}
\begin{equation} \label{eq1113}
Y \approx 1 + \Big[\frac{2 k_BT \tilde{\chi}^{SF}_{zz}}{L^3} \Big(1-\frac{2}{\pi}\Big) -\frac{2k_BT \tilde{\chi}^{AF}_{zz}} {L^3} \Big] / \Big(\tilde{m}^2_{z, \infty} - 2 \sqrt{\frac{2k_BT \tilde{\chi}^{SF}_{zz}}{{\pi}L^3}} \Big) 
\approx 1 \enskip ,
\end{equation}\\
\end{widetext}

\noindent and its height is

\begin{equation} \label{eq1114}
k_BT \tilde{\chi}^{', {\rm max}}_{zz} \cong \frac{L^3 \tilde{m}^2_{z, \infty}} {4} \Big(1 - 2 \sqrt{\frac{2k_BT \tilde{\chi}^{SF}_{zz}}{{\pi}L^3 \tilde{m}^2_{z, \infty}}} \Big) \enskip .
\end{equation}

\noindent This means that the leading correction is reduced by a factor of order $L^{-3/2}$. We also note that exactly at $H=H^t$ (i.e., $Y=q$),
$\tilde{\chi}'_{zz}/ \tilde{\chi}^{', {\rm max}}_{zz}= 4 q / (1 + q)^2 \approx 0.733$ (if we assume $q=\pi$). Estimation of this ratio offers yet another route to test the value of $q$.

%

The scaling function for $\tilde{\chi}_{xy}^{'}$ is given by

\begin{equation}
k_B T \tilde{\chi}_{xy}^{'} = L^3  (A Y^2 + BY + C)/(1 + Y)^2
\end{equation}
where the expressions $A, B, C$ are given (to the necessary order in inverse powers of $L$) by

\begin{equation}
A = k_B T \tilde{\chi}_{xy}^{SF} / L^3
\end{equation}

\begin{equation}
B = \psi_\infty^2 - 2\psi_\infty \sqrt{\pi k_BT \tilde{\chi}^{AF}_{xy}/(2L^3)}
\end{equation}

\begin{equation}
C = 2 k_B T \tilde{\chi}_{xy}^{AF} (1 - \pi/4) / L^3
\end{equation}
So from this expression one can see clearly the ``switching'' between the
two susceptiblities of the ``background'' phases, which result from this
expression when $Y = 0$ or $Y = \infty$, respectively, while 
\begin{equation}
Y = Y_{\max} = (B - 2C )/ (B - 2A) \approx 1+ (2A - 2C)/B
\end{equation}
yields the susceptibility maximum.

Thus, to leading order all susceptibility maxima occur at the same location, namely for

\begin{equation} \label{eq1116}
\mathcal{Z}=\mathcal{Z}_{\rm max} =-\ln q \enskip, \quad H^{\rm max} = H^t + \frac{k_BT \ln q}{ \Delta{m}L^3} \enskip ,
\end{equation}

\noindent but higher order corrections (of order $L^{-6})$ differ.  In both cases, the maximum staggered susceptibility also varies proportional to $L^3$ but has a $L^{-3/2}$ correction,

\begin{equation} \label{eq1117}
k_BT \tilde{\chi}^{', \rm max}_{xy} \cong L^3 \psi^2_\infty \frac{1}{4} \Big(1 - \sqrt{2 \pi k_BT \tilde{\chi}^{AF}_{xy} / (\psi^2_\infty L^3)}\Big) \enskip .
\end{equation}


\noindent and $\tilde{\chi}^{', {\rm max}}_{zz}$ is given by Eq.~(\ref{eq1114}). Here, too, $\tilde{\chi}'_{xy}$ for $H=H^t$ is smaller than $\tilde{\chi}^{', {\rm max}}_{xy}$ by the same factor $4q/(1 +q)^2$ as quoted above.

We now turn to the divergence of the susceptibility $\tilde{\chi}^{SF}_{xy}$ as $L \rightarrow \infty$:  Fisher and Privman ~\cite{FisherPrivman} predicted for isotropic, $n$-component magnets that $k_B T \tilde{\chi}_{xy}=\langle \psi^2 \rangle L^3=L^3 \psi^2_\infty/n + {\rm const}\;
L^2 (n-1)/(n+2)$ while Chen and Landau ~\cite{ChenLandau} predicted $\langle \psi^2 \rangle L^3=L^3 \psi^2_\infty/n + {\rm const}\;L^2 (n-1)/n$ (for $n=3$).
Thus, the leading correction is of order $1/L$ (rather than $L^{-3}$, as found in Eq.~(\ref{eq116})). Fisher and Privman predict that $L^3 (\langle \psi^2 \rangle -
\langle |\vec\psi| \rangle^2)$ varies proportional to $(n-1)L$ rather than being the finite constant obtained here.  It would be interesting to test these predictions (based on spin wave theory) using suitable numerical results for the present model, but this is a task that must be left for future work.

\section{}
Here we want to consider the effect of the nearby bicritical point on the finite size behavior of the 4th order cumulant of $\tilde{m}_z$ leading to the dramatic variation in crossing points in Fig.~\ref{cumMzs}.

\begin{figure}
\centering
\includegraphics[clip,angle=0,width=0.9\hsize]{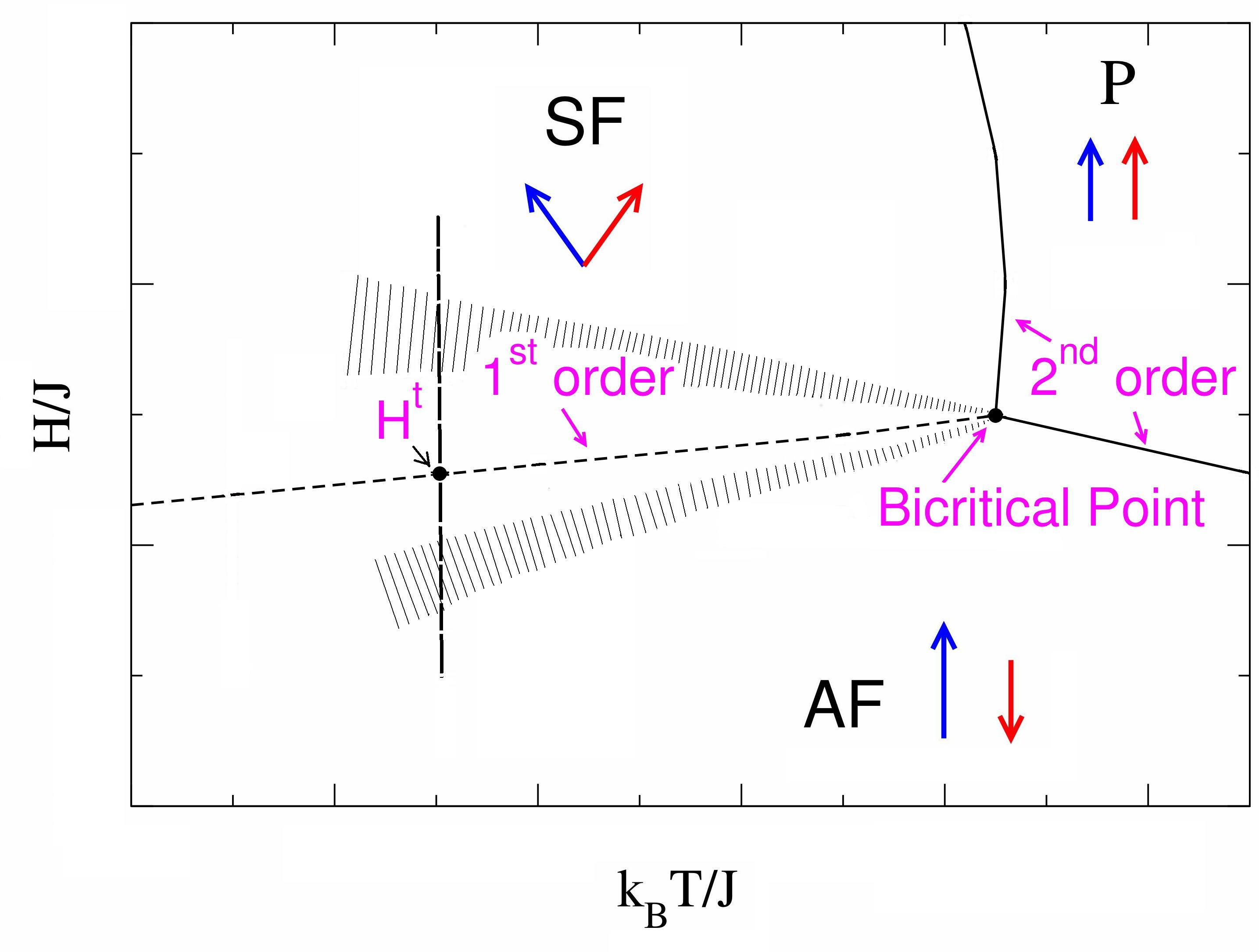}
\caption{\label{crossover} Phase diagram near the bicritical point showing crossover (shaded regions) between the bicritical and spin-flop regions.  (Crossover regions between the bicritical point and other phase boundaries are omitted for clarity.)}
\end{figure}

When the normal distance $d_{SF}(T,H)$ from the SF-P phase boundary inside the region of SF order is rather small,
the correlation length $\xi_{SF}$ of order parameter fluctuations is very large.  As is standard for critical phenomena, this is
described by the power law $\xi_{SF} \propto d^{-\nu_{xy}}_{SF}$, where $\nu_{xy}$ is the (universal) critical exponent of the XY model.
Likewise, in the region of the AF phase close to the AF-P phase boundary, the correlation length $\xi_{AF}$ of fluctuations of the AF order parameter
is very large, $\xi_{AF} \propto d^{-\nu_I}_{AF}$, $d_{AF} (T,H)$ being the normal distance from the AF-P phase boundary, and $\nu_I$ the
critical exponent of the Ising universality class. These power laws, however, apply only for a state point $(T,H)$ that is not too
close to the bicritical point $(T_b, H_b)$.  Close to the bicritical point, all components of the staggered magnetization are
simultaneously critical, and the fluctuations are characterized by Heisenberg criticality, with $\xi \propto d ^{- \nu_H}$, where $d$ is the
distance of the state point from the bicritical point. This latter relation applies when $d$ is sufficiently small and outside the
two shaded regions in the schematic sketch, Fig.~\ref{crossover}. Within the shaded regions, a smooth crossover to the first-order behavior at the spin-flop boundary 
occurs.\\

We conclude that near the AF-SF phase boundary for $T<T_b(H)$ there is a region where order parameter fluctuations
   of Heisenberg model type occur, and the correlation length $\xi$ of these fluctuations only gradually diminishes with the
   distance from the bicritical point. Even at the chosen temperature $T=0.95 J/k_B$, this correlation length must still be fairly large; and while we have not determined it directly, we can safely conclude this since probability distributions of energy (Fig.~\ref{probE}), magnetization (Fig.~\ref{probMz}), and order parameter components are still very broad for $L=40$.  Moreover, for smaller $L$ the two coexisting phases can hardly be recognized from these distributions.  The broadness of the peaks for $L=40$ in these figures is evidence that the corresponding (staggered) susceptibilities are still very large as well.\\

   From these observations we can also conclude that the behavior of various cumulants at $T=0.95J/k_B$ are still Heisenberg-like when $L \ll 40$, and  when $L \approx 40$ a gradual crossover from this critical behavior to the behavior characteristic for the first order transition begins to set in.\\

   To provide quantitative evidence for this scenario, we recall that for a Heisenberg antiferromagnet the order parameter distribution $P_L({\tilde{\vec{m}}})$ exhibits full rotational symmetry in order parameter space.  Using polar coordinates\\

   \begin{equation} \label{eqB1}
   {\tilde{\vec m}}_z = \tilde{m} \cos \theta, \, {\tilde{\vec m}}_x= \tilde{m} \sin \theta \cos \varphi, \, {\tilde{\vec m}}_y= \tilde{m} \sin \theta \sin \varphi
   \end{equation}

\noindent   we can write $P_L\Big({\tilde{\vec m}}\Big)  d {\tilde{\vec m}} =P_L (\tilde{m}) \tilde{m}^2 d \tilde{m} \sin \theta d \theta d \varphi$, where only a distribution $P_L (\tilde{m})$ of the magnitude $\tilde{m}$ of the order parameter is needed.

 The order parameter cumulant of the Heisenberg model

\begin{equation} \label{eqB2}
   U^H=1 - \Big \langle \Big({\tilde{\vec m}}^2 \Big)^2\Big\rangle / \Big( 3 \Big \langle {\tilde{\vec m}} ^2 \Big \rangle^2 \Big)
\end{equation}

\noindent at criticality is well known \cite{35,36}

\begin{equation} \label{eqB3}
U^H_*=0.620 (1).
\end{equation}

\noindent But Eq.~(\ref{eqB2}) is not what has been computed in the main text of this paper, where rather cumulants of ${\tilde{\vec m}}_z$ \{Eq.~(\ref{eq32VI})\} or the transverse order parameter $\vec{\psi}=({\tilde{\vec m}}_x, {\tilde{\vec m}}_y)$ \{Eq.~(\ref{eq32V})\} were considered. However, it turns out that it is straightforward to consider these quantities $U^z$, $U^{xy}$ for Heisenberg criticality as well, and actually both of them can be expressed in terms of $U^H$.\\
In order to see this, we first note that

\begin{equation} \label{eqB4}
\langle {\tilde{\vec m}}^2 \rangle = \int\limits^\infty_0 \tilde{m}^4 P_L (\tilde{m}) d \tilde{m}/ \int\limits^\infty_0 \tilde{m}^2 P_L (\tilde{m}) d \tilde{m} \quad,
\end{equation}

\begin{equation} \label{eqB5}
\langle {({\tilde{\vec m}}^2})^2 \rangle = \int\limits^\infty_0 \tilde{m}^6 P_L (\tilde{m}) d \tilde{m}/ \int\limits^\infty_0 \tilde{m}^2 P_L (\tilde{m}) d \tilde{m} \quad,
\end{equation}

\noindent the angular part simply cancels out in both expressions. Now from symmetry it is trivial to conclude that

\begin{equation} \label{eqB6}
\Big\langle {\tilde{\vec m}}^2_z \Big\rangle = \frac{1}{3} \Big\langle {\tilde{\vec m}}^2 \Big\rangle, \quad \Big\langle \psi^2 \Big\rangle = \frac{2}{3} \Big\langle {\tilde{\vec m}}^2 \Big\rangle
\end{equation}

\noindent while in $\langle \tilde{m}^4_z \rangle$, $\langle \psi^4 \rangle$ the angular parts contribute, but are straightforward to compute, e.g.

\begin{eqnarray} \label{eqB7}
&&\langle {\tilde{\vec m}}^4_z \rangle = \frac{\int\limits^\infty_0 \tilde{m}^6 P_L (\tilde{m}) d \tilde{m} \int\limits^\pi_0 \cos^4\theta \sin\theta d \theta}{ \int\limits_0^\infty \tilde{m}^2 P_L (\tilde{m}) d \tilde{m} \int\limits^\pi_0 \sin \theta d \theta} \nonumber\\
&&\qquad =\frac{1}{5} \Big\langle \Big({\tilde{\vec{m}}}^2\Big)^2 \Big\rangle
\end{eqnarray}

\noindent and similarly

\begin{equation} \label{eqB8}
\Big \langle \psi ^4 \Big \rangle =\frac{8}{15}\; \Big\langle \Big({\tilde{\vec{m}}}^2 \Big)^2 \Big\rangle \quad .
\end{equation}

Thus, we find

\begin{equation} \label{eqB9}
U^z =1 - \frac{3}{5} \frac{\langle({\tilde{\vec{m}}}^2 )^2 \rangle}{\langle {\tilde{\vec{m}}}^2\rangle^2} = \frac{9}{5} U_H - \frac{4}{5} \quad,
\end{equation}

\noindent and

\begin{equation} \label{eqB10}
U^{xy} =1 - \frac{2}{5} \frac{\langle({\tilde{\vec{m}}}^2 )^2 \rangle}{\langle {\tilde{\vec{m}}}^2\rangle^2} = \frac{6}{5} U_H - \frac{1}{5} \quad .
\end{equation}

Using Eq.~(\ref{eqB3}), we hence predict that in the bicritical region,

\begin{equation} \label{eqB11}
U^z_* \approx 0.316(2), \quad U^{xy}_* \approx 0.544(1) \quad.
\end{equation}
\\
 Looking at Fig.~\ref{cumMzs}, we see that for $L=40$, $L=50$, we still have spurious cumulant crossing somewhere in the region of $U^z_L \approx 0.34(3)$, while
for larger $L$ the crossings move towards significantly smaller values.  For the pair $L=(80,100)$ the crossing point is negative. We recall that
a degeneracy constant $q=\pi$ (see Eq.~(\ref{eq32VI})) implies $U^z_\infty =(2-q)/3=-0.3805$.  Thus a tentative interpretation of the behavior seen in Fig.~\ref{cumMzs} is a slow crossover from bicritical behavior to first-order-scaling. With respect to $U^{xy}_*$, Fig.~\ref{cumMxys} did not indicate massive crossover behavior; but this can be understood since $U^{xy}_*$ does not differ much from the first-order scaling prediction
 $U^{xy}_\infty \approx 0.56$.  Thus, the crossover between these values is ``masked'' by the standard corrections to finite size scaling.\\


\end{document}